\makeatletter
\def\@fnsymbol#1{\ensuremath{\ifcase#1\or **\or \dagger\or \ddagger\or
  \mathsection\or \mathparagraph\or \|\or \dagger\dagger\or
  \ddagger\ddagger\else\@ctrerr\fi}}
\makeatother
\documentclass[
superscriptaddress,
preprintnumbers,
nofootinbib,
nobibnotes,
notitlepage,
amsmath,amssymb,
aps,
twocolumn]{revtex4-1}

\usepackage[T1]{fontenc}
\usepackage{aas_macros}
\usepackage{graphicx}
\usepackage{dcolumn}
\usepackage{bm, color, colortbl}
\usepackage{lipsum, babel}
\usepackage{lineno}
\usepackage[citecolor=magenta]{hyperref}
\usepackage{orcidlink}
\hypersetup{colorlinks,linkcolor=magenta,citecolor=magenta,urlcolor=magenta}  
\usepackage{ctable}
\usepackage{array}
\usepackage{graphicx}
\usepackage[large]{subfigure}
\usepackage{amssymb, amsmath}
\usepackage[amssymb]{SIunits}
\usepackage[normalem]{ulem}
\usepackage{aas_macros}
\usepackage{natbib}

\usepackage{xcolor}
\usepackage{booktabs}
\usepackage{multirow} 
 
\newcolumntype{C}[1]{>{\centering\arraybackslash\hspace{0pt}}p{#1}}
\newcolumntype{B}{>{\columncolor{blue!30}[.5\tabcolsep]}C{2.6cm}}

\begin{document}

\newcommand{\Hunit}{{\rm km}\ {\rm s}^{-1}\rm{Mpc}^{-1}}
\newcommand{\Frank}[1]{\textcolor{blue}{(FJQ: #1)}}
\newcommand{\FrankEdit}[1]{\textcolor{red}{#1}} 
\newcommand{\blake}[1]{\textcolor{orange}{(blake: #1)}}

\newcommand{\hun}{\,\mathrm{km}\,\mathrm{s}^{-1}\mathrm{Mpc}^{-1}}
\newcommand{\es}[1]{\textcolor{blue}{(ES: #1)}}
\newcommand{\hanyue}[1]{\textcolor{pink}{(HW: #1)}}
\newcommand{\alex}[1]{\textcolor{red}{(AF: #1)}}
\newcommand{\fei}[1]{\textcolor{teal}{(FG: #1)}}
\newcommand{\KW}[1]{\textcolor{purple}{(KW: #1)}}

\title{Measuring Cosmic Neutrino Masses Independently of Dark Energy}

\makeatletter
\definecolor{fifagold}{RGB}{212,175,55}
\def\@fnsymbol#1{\ensuremath{\ifcase#1\or \text{\textcolor{fifagold}{\scalebox{0.75}{$\bigstar\bigstar$}}}\or \dagger\or \ddagger\or \mathsection\or \mathparagraph\or \|\or **\or \dagger\dagger\or \ddagger\ddagger\else\@ctrerr\fi}}
\makeatother
\author{Frank J. Qu\,\orcidlink{0000-0001-7805-1068}}
\email{jq247@cantab.ac.uk}
\affiliation{Kavli Institute for Particle Astrophysics and Cosmology, Stanford University, 452 Lomita Mall, Stanford, CA, 94305, USA}
\affiliation{Department of Physics, Stanford University, 382 Via Pueblo Mall, Stanford, CA, 94305, USA}

\author{Fei Ge\,\orcidlink{0000-0002-3833-8133}}
\affiliation{California Institute of Technology, 1200 E. California Blvd., Pasadena, CA, 91125, USA}
\affiliation{Kavli Institute for Particle Astrophysics and Cosmology, Stanford University, 452 Lomita Mall, Stanford, CA, 94305, USA}
\affiliation{Department of Physics, Stanford University, 382 Via Pueblo Mall, Stanford, CA, 94305, USA}

\author{Hanyue~Wang\,\orcidlink{0009-0009-3082-8021}}
\affiliation{Kavli Institute for Particle Astrophysics and Cosmology, Stanford University, 452 Lomita Mall, Stanford, CA, 94305, USA}
\affiliation{Department of Physics, Stanford University, 382 Via Pueblo Mall, Stanford, CA, 94305, USA}

\author{Emmanuel~Schaan\,\orcidlink{0000-0002-4619-8927}}
\affiliation{Kavli Institute for Particle Astrophysics and Cosmology, Stanford University, 452 Lomita Mall, Stanford, CA, 94305, USA}
\affiliation{SLAC National Accelerator Laboratory, 2575 Sand Hill Road, Menlo Park, California 94025, USA}

\author{W.~L.~Kimmy Wu\,\orcidlink{0000-0001-5411-6920}}
\affiliation{California Institute of Technology, 1200 E. California Blvd., Pasadena, CA, 91125, USA}
\affiliation{Kavli Institute for Particle Astrophysics and Cosmology, Stanford University, 452 Lomita Mall, Stanford, CA, 94305, USA}
\affiliation{SLAC National Accelerator Laboratory, 2575 Sand Hill Road, Menlo Park, California 94025, USA}

\author{Alexander~Friedland\,\orcidlink{0000-0002-5047-4680}}
\affiliation{SLAC National Accelerator Laboratory, 2575 Sand Hill Road, Menlo Park, California 94025, USA}

\author{Blake~D.~Sherwin}
\affiliation{DAMTP, Centre for Mathematical Sciences, University of Cambridge, Wilberforce Road, Cambridge CB3 0WA, UK}
\affiliation{Kavli Institute for Cosmology Cambridge, Madingley Road, Cambridge CB3 0HA, UK}

\date{\today}

\begin{abstract}

Neutrino oscillations establish that neutrinos are massive, providing the only laboratory detection of physics beyond the Standard Model. Direct kinematic experiments bound the electron-neutrino mass to $m_{\nu_e} < 0.45$~eV (KATRIN, 90\% CL), implying $\sum m_\nu \lesssim 1.3$~eV. Conversely, cosmology within $\Lambda$CDM is highly constraining: Planck CMB, CMB lensing, and DESI DR2 BAO yield $\sum m_\nu < 0.056$~eV (95\% CL), in 2--3$\sigma$ tension with the inverted-ordering floor (0.10~eV). However, this bound relies on $\Lambda$CDM, while data hint at an evolving dark energy.
To determine the model dependence of cosmic neutrino mass bounds, we deconstruct each probe's sensitivity to late-time physics and pursue two robust routes to a $\sum m_\nu$ bound:
(i) The existing dark-energy-marginalized route, retaining all data and marginalizing over $(w_0, w_a)$, is shown to also be immune to flexible binned and cubic w(a) histories, yielding $\sum m_\nu < 0.152$~eV, sharpening to $\sigma(\sum m_\nu) \approx 0.043$~eV with Simons Observatory lensing and Spec-S5 BAO.
(ii) A new late-Universe-free route combines primary CMB, marginalizing over acoustic-peak smoothing via $A_{\rm lens}$, with the reconstructed lensing spectrum $C_L^{\kappa\kappa}$, removing late-time expansion dependence by construction. 
This yields $\sum m_\nu < 0.41$~eV today, tightening to 0.31~eV (Simons Observatory) and 0.28~eV (cosmic-variance limit) across all tested dark-energy models.
These relaxed bounds trade statistical power for model independence. Interestingly, they land in the sensitivity range targeted by next-generation laboratory experiments like Project 8 ($m_{\nu_e} \sim 0.1$~eV), motivating vital synergies between future cosmological and terrestrial neutrino measurements.

\end{abstract}

\preprint{SLAC-PUB-260720}

\maketitle

\section{\label{sec:level1}Introduction}
The discovery of nonzero neutrino masses in neutrino oscillation experiments is one of the key breakthroughs in particle physics in the last forty years and the first laboratory measurement of physics beyond the Standard Model. 
An important outstanding problem is to establish the absolute values of the neutrino masses and the neutrino mass hierarchy (ordering). The type of the mass hierarchy has a dramatic impact on the predictions of the supernova neutrino signal, the prospects of the neutrinoless double-beta decay experiments, and  the models of flavor mixing in the leptonic sector. Oscillation experiments measure two mass-squared splittings,  $2.5 \times 10^{-3}$ eV$^2$\cite{T2K:2023smv,NOvA:2025tmb} and $7.5 \times 10^{-5}$ eV$^2$~\cite{KamLAND:2008dgz,JUNO:2025gmd}, implying a lower bound of 0.05 eV on the mass of the heavy state in the normal hierarchy case. In the case of the inverted hierarchy, this bound applies to two states, so that the sum of the masses is at least 0.10 eV. While recent results from JUNO~\cite{JUNONeutrino2026}, NOvA~\cite{NOvA:2025tmb} and T2K~\cite{T2KNeutrino2026} show some preference for the normal hierarchy, the fit is not without tensions, particularly as to the value of the CP-violating phase~\cite{T2K:2025wet}, and the issue is far from   settled. Conclusively establishing the type of mass hierarchy is one of the key goals of future measurements at JUNO and DUNE~\cite{DUNE:2020ypp}.

While there has been remarkable progress in direct experimental measurements of the absolute neutrino masses, the current laboratory bounds do not approach the lower bound set by the oscillation studies. The recent bound from the KATRIN experiment is $m_{\nu_e}< 0.45$~eV at 90\% confidence level, with 0.3 eV sensitivity expected in 2027, once the analysis of the 1000-day dataset is completed~\cite{KATRIN:2024cdt}. To go beyond that and reach the values of $m_{\nu_e} < 0.1$~eV, new experimental techniques are needed, such as cyclotron radiation emission spectroscopy in development for Project 8~\cite{Project8:2022wqh,Project8:2022hun}, which aims for eventual 0.04 eV electron neutrino mass sensitivity.

In view of this, cosmological measurements of the neutrino masses are extremely important. It must be stressed that the cosmological bounds constrains the neutrino mass sum $\sum m_\nu$. The joint analyses of CMB, BAO, and gravitational lensing reported constraints as strong as $\sum m_\nu < 0.056$ eV~\cite{desicollaboration2025desidr2resultsii,qu2025unifiedconsistentstructuregrowth,garciaquintero2025cosmologicalimplicationsdesidr2}, which, on the face of it, already disfavors the inverted mass hierarchy \cite{2026arXiv260618987J}. 
The tension reaches $\sim 2$--$3\sigma$ when the positivity prior is formally relaxed and an effective neutrino mass is allowed to take negative values~\cite{desicollaboration2025desidr2resultsii, Green:2024xbb, Lynch:2025ine, Craig:2024tky, 2026PhRvD.113d3514G, 2026JCAP...01..041C}. 
These findings, however, assume $\Lambda$CDM and are relaxed for extended dark energy models \cite{Elbers:2025vlz, 2026PDU....5202296F, 2026PhRvD.113h3503L}. 
The $\Lambda$CDM assumption is itself under pressure from the same data, which mildly prefer an evolving dark-energy equation of state~\cite{desicollaboration2025desidr2resultsii,garciaquintero2025cosmologicalimplicationsdesidr2,Shao:2024mag}.

Both inferences are further conditioned on the reionization
optical depth: the CMB-inferred $\tau$ is anticorrelated with
$\Omega_m$, so a larger optical depth, $\tau \simeq 0.09$, would
simultaneously relax the preferences for sub-minimal neutrino mass and
for evolving dark
energy~\cite{Sailer:2025lxj,jhaveri2026raisingreionizationopticaldepth,Sullivan:2026tas}.
Data combinations can in principle be constructed to measure neutrino masses without knowledge of the optical depth \cite{PhysRevD.107.123522}.
Throughout this work we adopt the standard low-$\ell$ constraint and
vary the dark-energy assumptions instead, noting that marginalizing
over the lensing-smearing amplitude $A_{\rm lens}$, an ingredient of
the late-Universe-robust channel constructed below, acts nearly
identically to removing the low-$\ell$ $\tau$
information~\cite{Weiner:2026sfm}.
The neutrino mass constraints are also loosened when curvature is allowed to vary \cite{2026arXiv260313208P, 2025JCAP...08..014C}.

These two behaviours, the low neutrino mass inferred within
$\Lambda$CDM and the preference for a dynamical dark-energy equation
of state, are not independent: $\sum m_\nu$ and the dark-energy
parameters $(w_0, w_a)$ are degenerate through the same low-redshift
distance measurements. The matter density $\omega_m$ inferred from
these late-time distances, combined with the CMB-calibrated sound
horizon, fixes the neutrino contribution as the offset between the
total matter density and that in cold dark matter and
baryons~\cite{Loverde:2024nfi, Lynch:2025ine}; the dark-energy equation of state
rescales precisely those late-time distances, so freeing it degrades
the $\sum m_\nu$ bound. This degeneracy is geometric in origin:
decomposing the $\sum m_\nu$ information into geometric and growth
components shows that the bias from a misspecified dark-energy model
enters through the geometric component, carried by the same
low-redshift distances, while the growth component is unaffected ~\citep{Ge:2026inprep}.

At the background level the interplay between BAO and CMB drives a preference of order $1\sigma$ to $2.3\sigma$ for negative effective neutrino masses, depending on the choice of CMB likelihood and whether two-point lensing information is retained~\cite{Green:2024xbb, Lynch:2025ine}. 
Freeing the dark-energy equation of state relaxes the bound substantially, because the dark energy density is degenerate with $\omega_m$ in the same low-redshift distances of $D_M(z)$ integral~\cite{desicollaboration2025desidr2resultsii}. 
A bound whose strength  rests on the late-time\footnote{We define ``late times'' as the redshift range $z < z_{\rm DE} = 0.706$, where dark energy contributes appreciably to $H(z)$; $z_{\rm DE}$ is the DESI LRG2 effective redshift, used both as the reference for the BAO distance differences (Section~\ref{sec:bao}) and as the cutoff in the $\alpha_{\rm DE}$ parametrization (Section~\ref{sec:unlensed}). The results are insensitive to this choice: varying $z_{\rm ref}$ across the other DESI tracer redshifts ($0.510 \le z_{\rm ref} \le 1.484$) shifts the late-Universe-free bound by less than $2\%$ on both mock and real data.} expansion therefore leans on precisely the part of the model currently in dispute. Conversely, a $\sum m_\nu$ constraint demonstrably insensitive to the dark energy model quantifies the neutrino-mass information the data carry independently of it, and indicates how much of the standard-$\Lambda$CDM
tension with the oscillation lower bound should be attributed to neutrino physics and how much to the dark energy assumption~\citep{PhysRevD.111.083535}.
Existing complementary work has pursued building data combinations independent of the sound horizon at recombination instead \cite{2026JCAP...02..034S}.

We take the combination apart probe by probe, asking how much $\sum m_\nu$ information each observable carries and how much of it is entangled with the assumed expansion history. 
Late-time data inform $\sum m_\nu$ almost entirely through their determination of the physical matter density $\omega_m$ (Appendix~\ref{sec:anatomy}, see also \cite{Lynch:2025ine}), and it is precisely this determination that dark-energy freedom degrades.
We explore the robustness of various neutrino mass bounds to extended dark energy models (see \cite{2025arXiv251208752N} for different data combinations).

We examine the
standard dark-energy-marginalized route in Section~\ref{sec:datasets}, the
unlensed primary CMB (Section~\ref{sec:unlensed}), the
two-point and four-point CMB lensing channels
(Section~\ref{sec:lensing}), and the BAO distances
(Section~\ref{sec:bao}) in turn, assemble the
pieces robust to dark energy's impact on the late Universe into the late-Universe-free combination and compare it with the
standard dark-energy-marginalized analysis (Section~\ref{sec:combined}).
Finally, we
close with prospects for both routes with future data
(Section~\ref{sec:forecast}).

\section{Datasets considered
\& Dark-energy-marginalized bound
}
\label{sec:datasets}

We employ three primary datasets. For the primary CMB (\textbf{CMB}),
We use the \texttt{CamSpec} high-ell temperature and polarization likelihood produced using \textit{Planck} PR4 maps ~\citep{Rosenberg:2022sdy}, as well as the \texttt{Commander} low-ell TT and \texttt{SRoll2} low-ell EE likelihoods made from PR3 maps \citep{prince2021pythoncompressedlowellplanck,Delouis:2019bub}. For CMB lensing (\textbf{CMBL}), we combine \textit{Planck} PR4, ACT DR6 (extended range), and SPT3G~\citep{ACT:2023dou,ACT:2023kun,ACT:2023ubw,Ge_2025,qu2025unifiedconsistentstructuregrowth}. For \textbf{BAO}, we use the DESI DR2 measurements~\citep{desicollaboration2025desidr2resultsi,desicollaboration2025desidr2resultsii}, which provide transverse ($D_M/r_d$) and radial ($D_H/r_d$) distance ratios (or isotropic $D_V/r_d$ for the lowest redshift bin) from seven tracer samples spanning $0.1 < z < 4.2$ (BGS, LRG1--3, ELG1--2, QSO, and Ly$\alpha$).

By convention, the widely used approach is to retain all the data and marginalize over a parametrized dark-energy history, typically $w(a) = w_0 + w_a(1-a)$, which tightens the bound to $\sum m_\nu < 0.152$~eV (consistent with \cite{Elbers:2025vlz}), a factor of $2.7$ in $\sigma(\sum m_\nu)$ below the parametrization-free result. 
A natural objection is that this marginalized bound depends on the chosen parametrization, and that a more flexible dark-energy model would weaken it further. We test this directly and find that, although it remains debatable whether $(w_0, w_a)$ is an adequate description of the dark-energy sector, it is effective at capturing the background expansion and growth to which $\sum m_\nu$ is sensitive: across progressively more flexible smooth expansion histories, from constant $w$ to $(w_0, w_a)$ to binned and cubic $w(a)$ parametrizations, the bound saturates near $0.152$~eV (Section~\ref{sec:combined}, Figure~\ref{fig:mnu_1d_data}), even though the data resolve three to four independent $w(a)$ modes (Section~\ref{sec:saturation}). Within smooth late-time expansion histories, $(w_0, w_a)$ marginalization is therefore sufficient. It nonetheless carries one caveat: it tightens with the same low-redshift data that drive the dark-energy preference, moving from $0.219$~eV on mock $\Lambda$CDM data used in Sect.~\ref{sec:strategy} to $0.152$~eV on the real data; its strength is therefore tied to those data rather than fixed by construction.

\section{Sensitivity of cosmological probes to late-time physics \& late-Universe-free combination}
\label{sec:strategy}
\label{sec:metric} 

Each cosmological observable encodes information about the neutrino mass sum through a combination of geometric distances and structure growth, and these two channels differ in their sensitivity to the low-redshift expansion history~\citep{Ge:2026inprep}. In this section, we examine the primary CMB, CMB lensing, and BAO individually, asking for each: how much neutrino-mass information does it carry, and how much of that information is contaminated by assumptions about dark energy? We assess this quantitatively and explain it intuitively for each probe.

A natural measure of dark-energy sensitivity is the ratio of the marginalized neutrino-mass uncertainty between a dark-energy-extended analysis and the $\Lambda$CDM analysis,
\begin{equation}
R \equiv \frac{\sigma(\sum m_\nu)_{\rm DE}}{\sigma(\sum m_\nu)_\Lambda},
\label{eq:R}
\end{equation}
so that $R \approx 1$ when freeing the dark-energy model leaves the bound intact and $R \gg 1$ when it widens it.\footnote{The parameter correlations from the chains support this interpretation of $R$: in the $(w_0, w_a)$ analysis of the standard combination the neutrino mass is strongly coupled to the equation of state, with correlation coefficients $r(\sum m_\nu, w_0) = +0.36$ and $r(\sum m_\nu, w_a) = -0.47$ on the real data (partial correlations $-0.59$ and $-0.65$ with the remaining cosmological parameters held fixed~\citep{Allali:2025yvp}), whereas for the late-Universe-free combination the couplings essentially vanish, $|r| \lesssim 0.1$; the neutrino-mass direction is nearly orthogonal to the dark-energy sector, and marginalizing over it is correspondingly free of cost.}
Unless stated otherwise, ``DE'' denotes the $(w_0, w_a)$ parametrization. For a bound to qualify as dark-energy independent we require $R \approx 1$ not for this single extension but across progressively more flexible parametrizations, from constant $w$ through $(w_0, w_a)$ to binned and cubic expansions of $w(a)$ (Section~\ref{sec:saturation}).

The ratio $R$ carries one important caveat that determines how we use it: $R \approx 1$ is necessary but not sufficient. A probe may return $R \approx 1$ not because its neutrino-mass information is insensitive to dark energy, but because it is not yet measured precisely enough for dark-energy freedom to act on.

For the same reason, $R$ is a property of a data combination rather than of its constituent probes: two datasets that individually return $R \approx 1$ can give $R > 1$ when combined, since the joint fit is sensitive to dark-energy-dependent information that neither dataset constrains on its own.

We therefore quote $R$ for each full combination we analyze rather than inferring its robustness from that of its parts, and we repeat the analysis at the sensitivity of upcoming experiments (Section~\ref{sec:forecast}), where channels that are prior-limited today become informative.

Two conventions for quoting a bound appear in this work and we clarify the convention here. Chain-based results, whether on real or
mock data, are quoted as one-sided $95\%$ credible limits, the 95th
percentile of the marginalized $\sum m_\nu$ posterior; this is the
quantity an analysis of the data reports. Fisher forecasts
(Figures~\ref{fig:forecast} and \ref{fig:lowz_cut}) are instead quoted
as the width $\sigma(\sum m_\nu)$. On mock data generated at
$\sum m_\nu = 0.06$~eV the two are related empirically by
${\rm UL} \simeq (3.3$--$3.7)\,\sigma(\sum m_\nu)$ across every
combination in this work: the posterior is truncated at
$\sum m_\nu \geq 0$ and positively skewed, so the credible limit
exceeds the Gaussian expectation $0.06~{\rm eV} + 1.645\,\sigma$ by up
to $\sim 50\%$ for the weakest combinations and approaches it for the
tightest. Comparisons between the two conventions, and with Fisher
forecasts in the literature, should therefore be made at the level of
$\sigma(\sum m_\nu)$, which we quote alongside the credible limits in
Section~\ref{sec:forecast}. The ratio $R$ is insensitive to the
choice, the relation between limit and width being common to numerator
and denominator.

Alongside $R$, which measures how much the posterior widens when the dark-energy family is freed, we report the change in best-fit $\chi^2$ between $\Lambda$CDM and each extended model, which quantifies whether the additional dark-energy freedom yields a statistically improved fit to the data: on the real data the standard combination prefers $(w_0, w_a)$ by $\Delta\chi^2 = -5.5$, with the gain concentrated in the BAO component, while the late-Universe-free combination shows no such preference for any model we test, $|\Delta\chi^2| \lesssim 2$ (Section~\ref{sec:combined}).

Throughout this section we evaluate $R$ on mock rather than real data, so that it reflects the intrinsic dark-energy response of each probe rather than statistical fluctuations, anomalies, or parameter pulls specific to the current datasets. We generate mock datasets for CMB lensing, BAO, and the primary CMB high-$\ell$ spectra, all adopting the \textit{Planck} NPIPE $\Lambda$CDM best-fit cosmology~\citep{Rosenberg_2022}; the low-$\ell$ TT (\texttt{Commander})~\citep{prince2021pythoncompressedlowellplanck} and low-$\ell$ EE (\texttt{SRoll2})~\citep{pagano/etal:2020} likelihoods are retained as the real \textit{Planck} data, so $\tau$ is fixed by the actual \texttt{SRoll2} measurement. This choice removes (i) the \textit{Planck} preference for $A_{\rm lens} > 1$~\citep{Rosenberg:2022sdy}, which artificially enhances the two-point lensing constraint on $\sum m_\nu$ (Appendix~\ref{app:fiducial_cmb}); and (ii) the mild discrepancy between the matter density $\Omega_m$ preferred by DESI DR2 BAO and by the \textit{Planck} CMB, which drives a preference for negative effective neutrino mass in their joint analysis~\citep{Lynch:2025ine,desicollaboration2025desidr2resultsii}. We present results with real data for all probes in Section~\ref{sec:combined}.

\subsection{\textbf{Unlensed CMB alone is robust to late-time physics}}
\label{sec:unlensed}

The primary CMB power spectrum is sensitive to the late-time expansion history through two main effects: the integrated Sachs-Wolfe (ISW) effect \citep{Hu_2002}, which modifies the large-scale temperature anisotropies, and the angular diameter distance to last scattering $D_A(z_*)$, that sets the projection of the acoustic scale onto the sky through $\theta_* = r_*/D_A(z_*)$. Both effects depend on the low-redshift expansion rate, and therefore on the dark energy model. A third and dominant source of late-time sensitivity in the observed CMB is gravitational lensing, which smooths the acoustic peaks and transfers power to small scales; we defer the discussion of lensing to Section~\ref{sec:lensing}.

To isolate the unlensed primary CMB, we marginalize over a lensing amplitude
parameter $A_{\rm lens}$ that rescales the lensing power spectrum
$C_L^{\phi\phi}$ where it enters the computation of the lensed two-point
spectra~\citep{Calabrese_2008} $C_\ell^{TT}$, $C_\ell^{TE}$, and
$C_\ell^{EE}$; it leaves the ISW contributions to the unlensed anisotropies
untouched. The late-time ISW effect couples to $\sum m_\nu$ through the shift
in the epoch of dark-energy domination at fixed $\theta_\star$; it is
therefore the only channel in the unlensed spectra through which dark energy
carries neutrino-mass information. That channel is confined to multipoles
limited by cosmic variance, and we verify below that excising the low-$\ell$
TT data that carry it changes neither the bound nor $R$. With $A_{\rm lens}$
free, the remaining neutrino-mass information comes from the early ISW effect and the geometric distance $D_A(z_\star)$.

The early-ISW origin follows from the timing of the non-relativistic
transition. With mean momentum $\langle p_\nu \rangle \approx
3.15\,T_\nu$, a neutrino of mass $m_\nu$ becomes non-relativistic at
$z_{\rm nr} \approx 1890\,(m_\nu/{\rm eV})$, after photon decoupling
for $m_\nu < 0.58$~eV and hence for every mass in the range considered
here. The acoustic oscillations and their radiation-driving envelope
are therefore set while the neutrinos are still relativistic, and the
perturbation-level imprint arises from the modified decay of the
gravitational potentials after decoupling\footnote{We verified this
decomposition directly: switching off the early-ISW source term in a
Boltzmann calculation removes two-thirds to three-quarters of the
$\sum m_\nu$ response of the unlensed $C_\ell^{TT}$ over
$20 \le \ell \le 500$ at fixed $(\omega_b, \omega_{\rm cdm},
\theta_\star)$, with much of the remainder attributable to the
late-time ISW term at $\ell \lesssim 50$; the residual response at
higher multipoles is below the $0.1\%$ level.}, imprinted at
$50 \lesssim \ell \lesssim 200$
\citep{LESGOURGUES_2006, Archidiacono_2017}. 

\begin{figure}[t]
    \centering
    \includegraphics[width=\columnwidth]{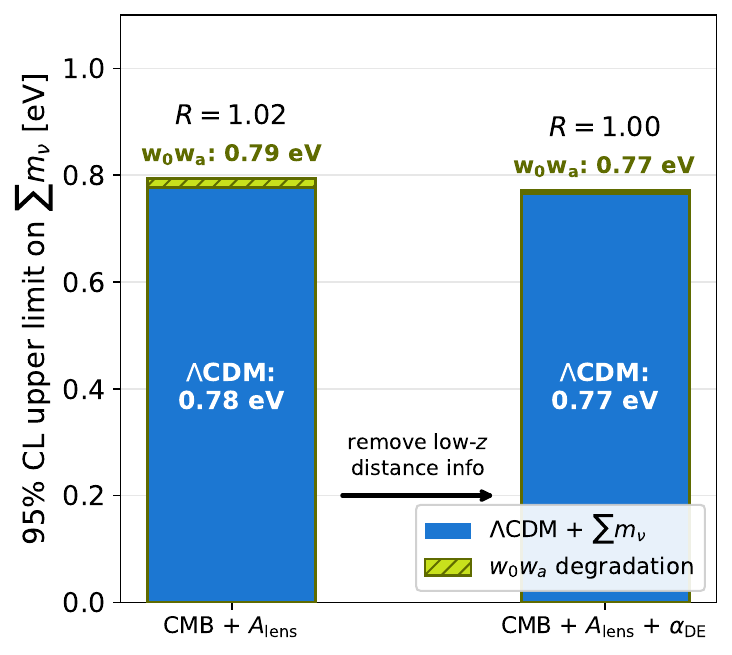}
    \caption{$95\%$ upper limits on $\sum m_\nu$ from the unlensed primary CMB, obtained by marginalizing over the lensing amplitude $A_{\rm lens}$. Blue bars show the $\Lambda$CDM constraint; the hatched green region indicates the degradation when dark energy parameters $(w_0, w_a)$ are freed. The ratio $R \equiv \sigma_{w_0 w_a}/\sigma_\Lambda$ quantifies the sensitivity to the dark energy model. With $A_{\rm lens}$ alone (left), the constraint is $\sim 0.78$~eV with $R = 1.02$, indicating negligible dark energy sensitivity. Adding $\alpha_{\rm DE}$ marginalization (right), which removes the residual low-redshift distance information through $D_A(z_*)$, produces no further change ($R = 1.00$); the $\sim 0.01$~eV shift between configurations is consistent with MCMC sampling noise.}
    \label{fig:unlensed_cmb}
\end{figure}

As shown in Fig.~\ref{fig:unlensed_cmb}, the unlensed CMB alone
provides only weak constraints: the $95\%$ upper limit on $\sum m_\nu$
is $0.78$~eV in $\Lambda$CDM and $0.79$~eV in $w_0 w_a$CDM, with
$R = 1.02$. The near-unity ratio confirms that the unlensed CMB alone
is largely insensitive to dark energy assumptions; the ISW effect and
the geometric degeneracy contribute negligible dark energy-dependent
information about $\sum m_\nu$\footnote{We verified this directly by
repeating the analysis without the low-$\ell$ TT data, which carry the
bulk of the late-time ISW signal, and find a negligible change in the
bound and in $R$.}.

To further verify that any residual geometric sensitivity of $\sum m_\nu$ through $D_A(z_*)$ is small, we introduce a phenomenological parameter $\alpha_{\rm DE}$ that marginalizes over the late-time distance by rescaling $H(z)$ at low redshift during the Boltzmann integration:
\begin{equation}
H(z) \to
\begin{cases}
\alpha_{\rm DE} \times H_{\rm template}(z), & z < z_{\rm DE} \\
H(z), & z \geq z_{\rm DE}
\end{cases}
\label{eq:alpha_DE}
\end{equation}
where $H_{\rm template}(z)$ is the \textit{Planck} $\Lambda$CDM best-fit expansion history and $z_{\rm DE} = 0.706$. A single rescaling parameter is sufficient because the CMB acoustic peak positions depend on the integrated distance $D_A(z_*) \propto \int dz/H(z)$, not on the detailed shape of $H(z)$ at low redshift (see Appendix~\ref{app:alpha_DE} for implementation details)~\footnote{Note that $\alpha_{\rm DE}$ rescales only the late-time ($z < z_{\rm DE}$)
expansion rate; the comoving sound horizon $r_\star$ is fixed by pre-recombination
physics (set by $\omega_b$ and $\omega_c$) and is left untouched. Because the acoustic
scale $\theta_\star = r_\star/D_A(z_\star)$ enters through the sampled parameter
$\theta_{\rm MC}$ and is pinned by the data, the fit reabsorbs the $\alpha_{\rm DE}$
shift in $D_A(z_\star)$ into the inferred $H_0$ rather than displacing the acoustic
peaks; $\alpha_{\rm DE}$ thus marginalizes the low-redshift contribution to
$D_A(z_\star)$ along the $\theta_\star$--$H_0$ degeneracy while holding $r_\star$ fixed.}.
Adding $\alpha_{\rm DE}$ marginalization on top of $A_{\rm lens}$ yields $R = 1.00$, with no degradation in $\sigma(\sum m_\nu)$ relative to the $A_{\rm lens}$-only case (Fig.~\ref{fig:unlensed_cmb}, right). $R \approx 1$ in the $A_{\rm lens}$-only test already shows that the primary CMB's residual $\sum m_\nu$ sensitivity is itself dark-energy-insensitive; the absence of further degradation under $\alpha_{\rm DE}$ marginalization confirms that the late-time distance channel $D_A(z_*)$ carries negligible $\sum m_\nu$ information in the primary CMB.

These results demonstrate that the unlensed primary CMB provides a robust upper bound $\sum m_\nu \lesssim 0.8$~eV from early-universe physics alone. This bound is set predominantly by recombination-era physics and is therefore insensitive to assumptions about the late-time expansion history; while it does not reach the $\sim 0.1$~eV regime relevant for distinguishing the oscillation mass hierarchies, it confirms that the absolute neutrino mass scale lies at most at the eV level from CMB physics alone \citep{Ichikawa_2005}. 
For comparison, the standard CMB analysis without $A_{\rm lens}$ marginalization yields $0.57$~eV in $\Lambda$CDM on mock data; the $\sim 40\%$ degradation when marginalizing over $A_{\rm lens}$ shows that a substantial fraction of the neutrino-mass information in the CMB two-point power spectra originates from gravitational lensing, not from the unlensed acoustic peaks. We examine this lensing information in detail in the next section.

\subsection{Contrasting two-point and four-point lensing}
\label{sec:lensing}

\subsubsection{Four-point lensing alone is insensitive to late-time physics; Two-point lensing is not}

Gravitational lensing of the CMB encodes information about the neutrino mass sum through two distinct channels. The first is the two-point channel: lensing smooths the acoustic peaks and generates power to small scales in the observed $C_\ell^{TT}$, $C_\ell^{TE}$, and $C_\ell^{EE}$ power spectra \citep{LEWIS_2006}. This effect is captured in the standard (lensed) CMB analysis and corresponds to the information that is removed when $A_{\rm lens}$ is marginalized over. The second is the four-point channel: the lensing-induced non-Gaussianity in the CMB temperature and polarization fields can be exploited through quadratic estimators to reconstruct the lensing convergence power spectrum $C_L^{\kappa\kappa}$, which provides a direct measurement of the projected matter power spectrum \citep{Hu_2002,Okamoto_2003}.

\begin{figure*}[t]
    \centering
    \includegraphics[width=0.8\textwidth]{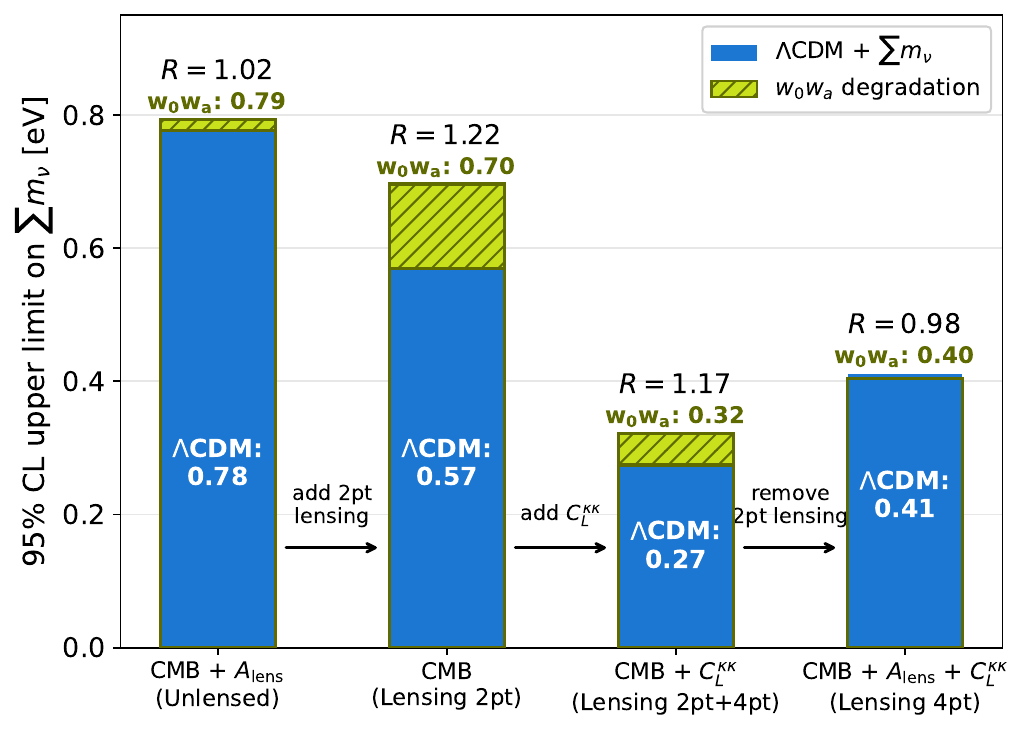}
    \caption{$95\%$ upper limits on $\sum m_\nu$ illustrating the constraining power and dark energy sensitivity of each CMB lensing channel, using mock primary CMB spectra ($A_{\rm lens} = 1$) and mock lensing data. Blue bars show the $\Lambda$CDM constraint; the hatched green region indicates the additional degradation when dark energy parameters $(w_0, w_a)$ are freed. The ratio $R \equiv \sigma_{w_0 w_a}/\sigma_\Lambda$ quantifies sensitivity to the dark energy model. Starting from the unlensed CMB (first bar; $A_{\rm lens}$ marginalized, $R = 1.02$), restoring the two-point lensing information tightens the constraint but introduces dark energy sensitivity ($R = 1.22$; second bar). Adding the reconstructed four-point lensing power spectrum $C_L^{\kappa\kappa}$ improves the constraint further ($R = 1.17$; third bar). Isolating the four-point channel by marginalizing over $A_{\rm lens}$ while retaining $C_L^{\kappa\kappa}$ (fourth bar) yields a constraint ($0.41$~eV) that is now tighter than the two-point channel alone ($0.57$~eV) and entirely insensitive to dark energy ($R = 0.98$). The reversal of the two-point and four-point hierarchy compared to the real-data version of this figure (Figure~\ref{fig:lensed_cmb_real}) confirms that the apparent strength of the two-point channel in real data is enhanced by the well-known \textit{Planck} preference for $A_{\rm lens} > 1$.}
    \label{fig:lensed_cmb}
\end{figure*}

Fig.~\ref{fig:lensed_cmb} illustrates the constraining power and dark energy sensitivity of each channel. Starting from the unlensed CMB ($\sum m_\nu < 0.78$~eV, $R = 1.02$; first bar), restoring the two-point lensing information by fixing $A_{\rm lens} = 1$ tightens the constraint to $0.57$~eV in $\Lambda$CDM (second bar).
However, the two-point channel also introduces significant dark energy sensitivity: the $w_0 w_a$ constraint degrades to $0.70$~eV, yielding $R = 1.22$.

Adding the reconstructed four-point lensing power spectrum $C_L^{\kappa\kappa}$ (third bar) tightens the constraint further to $0.27$~eV ($\Lambda$CDM) and $0.32$~eV ($w_0 w_a$CDM), with $R = 1.17$. Within the CMB-only analysis (no BAO), restoring both lensing channels tightens the bound by a factor of approximately three over the unlensed CMB ($0.78 \to 0.27$~eV in $\Lambda$CDM). This factor-of-three contribution from CMB lensing is less commonly emphasized in analyses of the standard $\sum m_\nu$ bound, which typically focus on the CMB + BAO combination. The degradation ratio is essentially unchanged from the two-point-only case, indicating that the four-point channel does not introduce additional dark energy sensitivity beyond what is already present in the two-point spectra.

To isolate the four-point channel, we marginalize over $A_{\rm lens}$ while including $C_L^{\kappa\kappa}$ (fourth bar). The constraint is $0.41$~eV in $\Lambda$CDM and $0.40$~eV in $w_0 w_a$CDM, with $R = 0.98$. The four-point lensing channel alone therefore provides a competitive neutrino-mass constraint that is insensitive to the dark-energy model, and on mock data it is in fact tighter than the two-point $\Lambda$CDM constraint ($0.41$~eV versus $0.57$~eV). When the real \textit{Planck} spectra are used instead (Appendix~\ref{app:fiducial_cmb}), the two-point constraint tightens to $0.36$~eV, appearing stronger than the four-point channel; this reversal is driven by the \textit{Planck} NPIPE preference for $A_{\rm lens} = 1.095 \pm 0.056$~\citep{Rosenberg_2022}, which enhances the sensitivity of the lensed CMB power spectra to $\sum m_\nu$. The four-point constraint is unaffected by this fluctuation, as $A_{\rm lens}$ marginalizes over the two-point contribution. We note, however, that even relative to an $A_{\rm 2pt}$-marginalized cosmology, the four-point lensing amplitude itself remains mildly high \citep{Ge_2025}.

\subsubsection{Physical origin of the different dark energy sensitivities}
\label{sec:lensing_de_sensitivity}

The preceding results show that the four-point lensing channel is less sensitive to the dark energy model while the two-point channel is not. Both channels probe the same underlying lensing power spectrum $C_L^{\kappa\kappa}$, but they weigh it differently as a function of lensing multipole $L$, and thus source redshift $z$ \citep{Smith_2006, Benoit_L_vy_2012}.

The four-point channel measures $C_L^{\kappa\kappa}$ directly at each multipole $L$ via
quadratic estimators \citep{Hu_2002}; its Fisher information peaks at $L \approx 300$, set by the noise
weighting of the combined CMB lensing reconstruction. 

\begin{figure}
\centering
\includegraphics[width=\columnwidth]{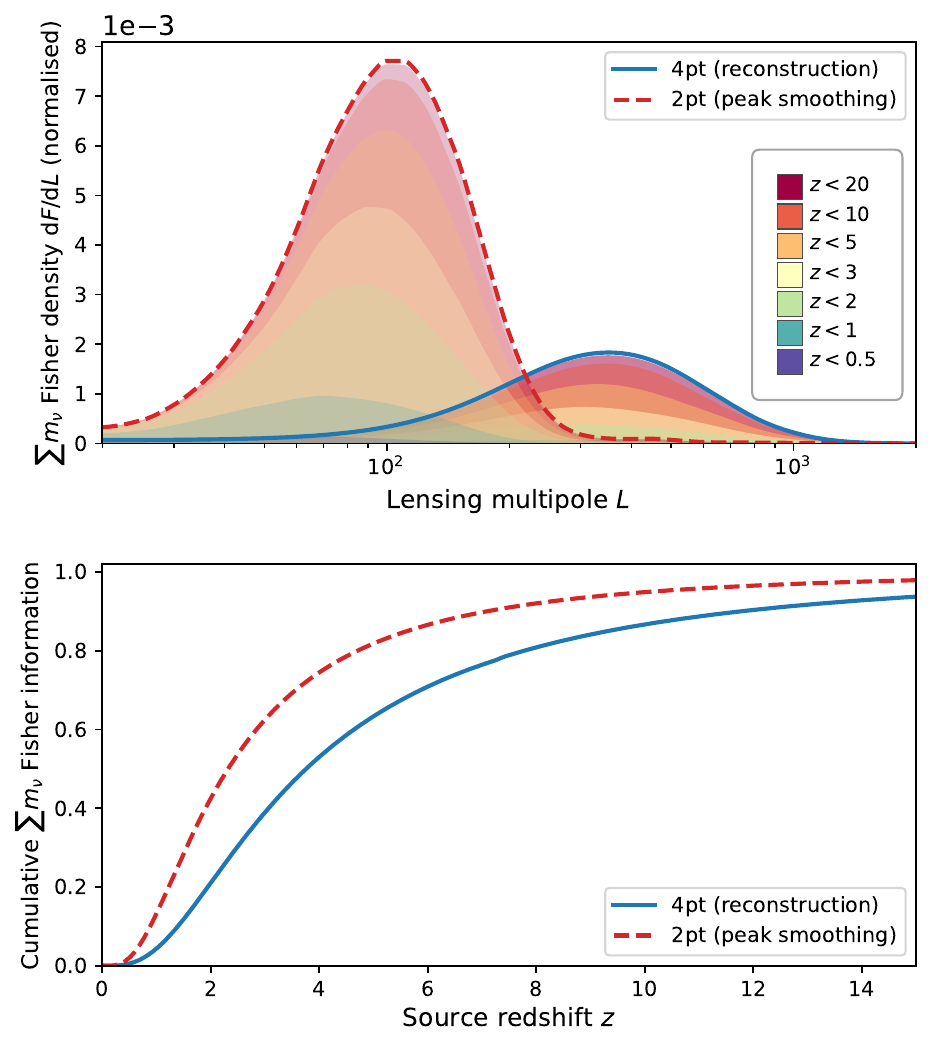}
\caption{
Decomposition of the $\sum m_\nu$ Fisher information carried by the
two-point (peak smoothing, dashed red) and four-point (reconstruction,
solid blue) lensing channels. We plot the noise-weighted Fisher
information density, $\mathrm{d}F/\mathrm{d}L \propto (\partial
C_L^{\phi\phi}/\partial\!\sum m_\nu)^2/\sigma_L^2$; the two
peak at different multipoles because the inverse-variance weighting
and the squaring shift the constraining power toward higher $L$ than
the derivative alone (which peaks near $L\sim100$; cf.\
\citep{Loverde_2024}). The four-point Fisher
density is weighted by the combined ACT DR6 + SPT + \textit{Planck}
lensing power spectrum uncertainties; the two-point Fisher density is
weighted by the \textit{Planck} CMB temperature power spectrum
uncertainties. \textbf{Top:} Fisher information density per lensing
multipole $L$, normalised to unit area, with stacked fills showing the
cumulative contribution from source redshifts $z < z_{\rm cut}$. The
four-point channel peaks at $L \approx 300$, while the two-point
channel peaks at $L \approx 100$. \textbf{Bottom:} cumulative fraction
of the total Fisher information as a function of source redshift. The
two-point channel draws $14\%$ of its information from $z < 1$ (median
source redshift $z_{1/2} \approx 2.4$); the four-point channel draws
only $4\%$ from $z < 1$ ($z_{1/2} \approx 3.8$). Both effects compound:
the two-point channel weights lower lensing multipoles, and at those
multipoles a larger fraction of the signal originates from the
dark-energy era. The two-point redshift weighting is a geometric
property of the lensing convolution and is insensitive to the
experimental noise model (see Appendix~\ref{app:2pt_noise_robustness}).
}
\label{fig:2pt_4pt_decomposition}
\end{figure}

The two-point channel enters indirectly, through the smoothing of the
acoustic peaks, and two scales select the lensing multipoles
responsible. The deflection variance is dominated by large lenses
($[L(L+1)]^2 C_L^{\phi\phi}/2\pi$ peaks at $L \approx 40$, with
$80\%$ of the variance accumulated by $L = 100$~\citep{LEWIS_2006}),
but a deflection field coherent over scales much larger than the
acoustic peak spacing $\Delta\ell \approx 300$ translates the spectrum
without distorting it, so the smoothing efficiency of a lens rises
with $L$\footnote{Beyond the damping tail, $\ell \gtrsim 3000$, the
lensed spectrum instead approaches the deflection power weighted by
the mean-squared unlensed CMB gradient~\citep{LEWIS_2006}. This regime
lies above the range of the \textit{Planck} spectra used here
($\ell \le 2500$); ground-based temperature spectra extend further,
but extragalactic foregrounds dominate the temperature power at these
multipoles, and we do not use them in our combination.}. Falling
deflection power and rising efficiency together select intermediate
multipoles: the response kernel
$K_{\ell L} \equiv \partial \tilde{C}_\ell^{TT} / \partial C_L^{\phi\phi}$,
computed numerically in Appendix~\ref{app:response_kernel}, shows the
two-point Fisher information peaking at $L \approx 100$, in agreement
with the principal-component analysis of \citep{Smith_2006}, whose
leading eigenmode from the lensed $\{T,E\}$ spectra has median
multipole $L = 114$, and well below the four-point case
(Figure~\ref{fig:2pt_4pt_decomposition}, top panel).

Two effects drive the asymmetry of the $\Sigma m_\nu$ constraint response
to dark energy models from 2pt and 4pt lensing. Neutrino free-streaming
suppresses structure on small scales, so the fractional effect of
$\Sigma m_\nu$ on $C_L^{\kappa\kappa}$ (the logarithmic derivative
$\partial \ln C_L^{\kappa\kappa}/\partial \Sigma m_\nu$ at the fiducial
cosmology) is small at low $L$, rises steeply through $L \sim 100$, and
largely saturates above $L \sim 300$ \citep{Loverde_2024}. The four-point
Fisher density therefore peaks at $L \approx 300$ not because the response
itself peaks there, but because the bandpower signal-to-noise of the
combined reconstruction is maximal at $L \approx 300$, where the response
is already close to its asymptotic value. Dark energy operates only at low
redshift; through the low-$z$ portion of the lensing kernel, its fractional
effect on $C_L^{\kappa\kappa}$ relative to that of $\Sigma m_\nu$ is
largest at low $L$, where the kernel draws more weight from $z<1$: the
ratio of the $w$ response to the $\Sigma m_\nu$ response falls from
$\simeq 5$ at $L = 10$ to unity near $L \simeq 100$ and $\simeq 0.9$ above
$L \sim 300$. The two-point channel, with Fisher information peaking near
$L \sim 100$ (source redshift $z\approx 1.3$, $14\%$ of its
$\Sigma m_\nu$ sensitivity from $z<1$), therefore lies where these two
effects together make $\Sigma m_\nu$ and dark energy most degenerate,
while the four-point channel, peaking near $L \sim 300$ ($z\approx 2.2$,
only $4\%$ from $z<1$; Figure~\ref{fig:2pt_4pt_decomposition}, bottom
panel), lies where the $\Sigma m_\nu$ response is saturated, the relative
dark energy response is weak, and the two parameters cleanly separate.
This is why marginalizing over $A_{\rm lens}$ to remove the two-point
channel yields a constraint with $R\approx 1$, independent of the
experimental noise (Appendix~\ref{app:2pt_noise_robustness}).

The low-redshift fraction of the $\sum m_\nu$ information carried by
the four-point channel is set by the lensing-kernel geometry rather
than by the experiment, rising only from $4.4\%$ ($z<1$) at current
noise to $7.7\%$ at the cosmic-variance limit; the dark-energy
insensitivity established here is therefore expected to persist as
the reconstruction improves. We confirm this expectation at forecast
precision in Section~\ref{sec:forecast}.

\subsection{BAO constrains $\sum m_\nu$ primarily through late-time-sensitive channels}
\label{sec:bao}

BAO measurements provide precise geometric constraints on the
expansion history through the transverse comoving angular diameter distance
$D_M(z)/r_d$ and the Hubble distance $D_H(z)/r_d \equiv c/[H(z)\, r_d]$
at multiple redshifts. Combined with the CMB, which calibrates the
sound horizon $r_d$ and pins the angular acoustic scale $\theta_*$,
BAO breaks the CMB geometric degeneracy and tightens the joint
constraint on $\Omega_m$ and $H_0$. Since $\omega_m = \omega_{cb} +
\omega_\nu$ with $\omega_\nu = \sum m_\nu / 93.14$~eV~\citep{LESGOURGUES_2006},
this in turn propagates into a constraint on $\sum m_\nu$
\citep{Lynch:2025ine, Loverde_2024}. However, BAO distances
also carry significant sensitivity to dark energy, since
$D_M(z) = \int_0^z c\,dz'/H(z')$ is a cumulative integral over the
full expansion history: even measurements at high redshift encode
$H(z')$ at low redshift where dark energy dominates. Allowing
$(w_0, w_a)$ freedom changes the low-redshift contribution to this
integral, degrading the $(\Omega_m, H_0)$ constraint and propagating
into $\sum m_\nu$.

To isolate the high-redshift geometric information from the 
transverse distances $D_M(z)/r_d$, 
we may construct differences
\begin{equation}
\Delta D_M(z)/r_d \equiv D_M(z)/r_d - D_M(z_{\rm ref})/r_d 
= \frac{1}{r_d}\int_{z_{\rm ref}}^{z} \frac{c\,dz'}{H(z')}
\end{equation}
relative to a reference redshift $z_{\rm ref} = 0.706$, retaining only tracers at $z > z_{\rm ref}$ and 
discarding all BAO measurements at lower redshift, including the 
radial distances $D_H(z)/r_d$, which we consider separately below. 
Crucially, both terms in the difference are measured distances from the data, so that no modeling is required.
The differencing removes the integral from $z = 0$ to $z_{\rm ref}$, 
where dark energy contributes most strongly to $H(z)$,
making it insensitive to the dark-energy model.
However, unfortunately,
we now show that it carries essentially no constraining 
power on $\sum m_\nu$ beyond what the CMB + lensing baseline already 
provides.

First, this differencing degrades the constraining power. In the matter-dominated regime, $D_M(z) \propto 1/\sqrt{\Omega_m}$ at fixed $H_0$, so all high-redshift transverse distances share a common $\Omega_m$-dependent scaling. The difference $\Delta D_M$ partially cancels this scaling, reducing $\partial(\Delta D_M/r_d)/\partial\Omega_m$ relative to $\partial(D_M/r_d)/\partial\Omega_m$; the cancellation is strongest for tracers closest to $z_{\rm ref}$. A Fisher analysis using DESI DR2 errors shows that the differenced observable yields $\sigma(\omega_m)$ approximately $2\times$ larger than absolute distances over the same redshift range (Appendix~\ref{sec:bao_diff_omegam}). Additional degradation arises from error propagation: since all differences share the same reference measurement, the covariance matrix acquires off-diagonal elements $\mathrm{Cov}(\Delta D_{M,i}, \Delta D_{M,j}) = \sigma^2_{D_M}(z_{\rm ref})$ that further reduce the original BAO constraining power.

\begin{figure}
\centering
\includegraphics[width=\columnwidth]{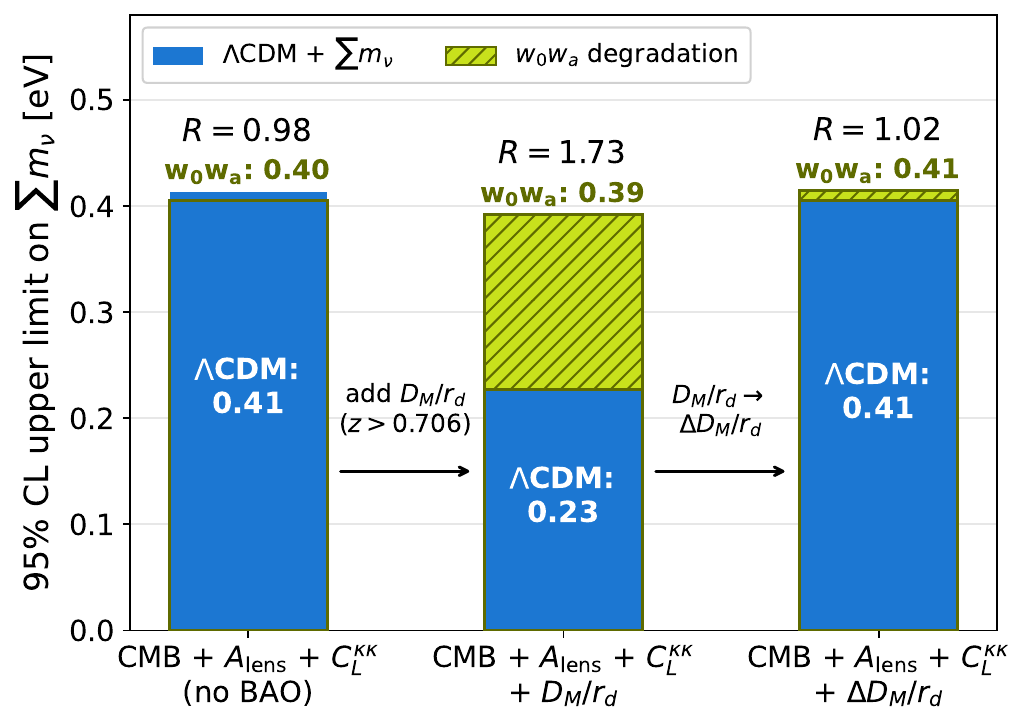}
\caption{Isolating the effect of distance differencing on transverse
BAO constraints. All three configurations use the DE-insensitive
CMB baseline (CMB + $A_{\rm lens}$ + $C_L^{\kappa\kappa}$) and
include no radial distance ($D_H/r_d$) information. Adding absolute
$D_M/r_d$ measurements at $z > 0.706$ (middle bar) tightens the
constraint from 0.41 to 0.23~eV in $\Lambda$CDM but introduces
strong dark energy sensitivity ($R = 1.73$). Replacing absolute
distances with differences $\Delta D_M/r_d$ relative to
$z_{\rm ref} = 0.706$ (right bar) restores $R = 1.02$, with the
constraint returning to 0.41~eV. The $w_0 w_a$ bound is unchanged
across all three configurations ($0.39$--$0.41$~eV), a consistency
check that the added transverse distances carry no $\sum m_\nu$
information once the dark-energy model is freed; the degradation
$R = 1.73$ in the middle configuration therefore reflects only the
$\Lambda$CDM-side tightening from absolute $D_M/r_d$, which the
$(w_0, w_a)$ freedom fully absorbs through the $D_M(z)$ integral.
All constraints use mock data.}
\label{fig:bao_bar}
\end{figure}

We now examine the impact of BAO on the DE-insensitive combination 
established in Section~\ref{sec:lensing}. Starting from the CMB + 
$A_{\rm lens}$ + $C_L^{\kappa\kappa}$ baseline, which isolates the 
four-point lensing channel and yields $\sum m_\nu < 0.41$~eV with 
$R = 0.98$, we add BAO information in three configurations 
(Figure~\ref{fig:bao_bar}).

Adding absolute transverse distances $D_M/r_d$ at $z > z_{\rm ref}$
tightens the constraint to $\sum m_\nu < 0.23$~eV, but reintroduces
dark energy sensitivity: $R$ rises to 1.73. This is expected: even
though $A_{\rm lens}$ has removed the DE-sensitive two-point lensing
channel from the CMB, the cumulative integral
$D_M(z) = \int_0^z c\, dz'/H(z')$ encodes the full low-redshift
expansion history where dark energy is dynamically important.

Replacing the absolute transverse distances with differences
$\Delta D_M/r_d$ relative to $z_{\rm ref}$ restores dark energy
independence ($R = 1.02$), but the constraint returns to
$\sum m_\nu < 0.41$~eV (Figure~\ref{fig:bao_bar}). The differenced
transverse distances do tighten $\sigma(\Omega_m)$ and $\sigma(H_0)$
once $(w_0, w_a)$ are marginalised, but this tightening is along the
$\omega_m = \Omega_m h^2 = $ const direction in the $(\Omega_m, H_0)$
plane, which is orthogonal to the $\omega_m$ direction that the CMB
acoustic peak ratios and damping tail already constrain. Since the neutrino mass enters through 
$\omega_\nu = \omega_m - \omega_b - \omega_c$, with the CMB pinning 
$\omega_b + \omega_c$ from the acoustic 
peak heights and damping tail~\citep{Lynch:2025ine,Loverde_2024},
the geometric tightening that $\Delta D_M/r_d$ provides does not
propagate into $\sigma(\sum m_\nu)$. The differenced transverse
distances therefore add negligible constraining power on $\sum m_\nu$
beyond what is already provided by four-point lensing; the neutrino
mass information in this combination is driven almost entirely by the
reconstructed lensing power spectrum.\footnote{We verify this directly on mock data by comparing the chains 
$\mathrm{CMB} + A_{\rm lens} + C_L^{\kappa\kappa}$ with and without 
$\Delta D_M/r_d$. In $\Lambda$CDM, where the CMB and lensing already fix 
$\Omega_m$ and $H_0$ through $\omega_m$, the marginal posteriors of all 
cosmological parameters are unchanged within MCMC sampling noise, with 
$\sum m_\nu < 0.41$~eV in both cases. }

\begin{figure}
\centering
\includegraphics[width=\columnwidth]{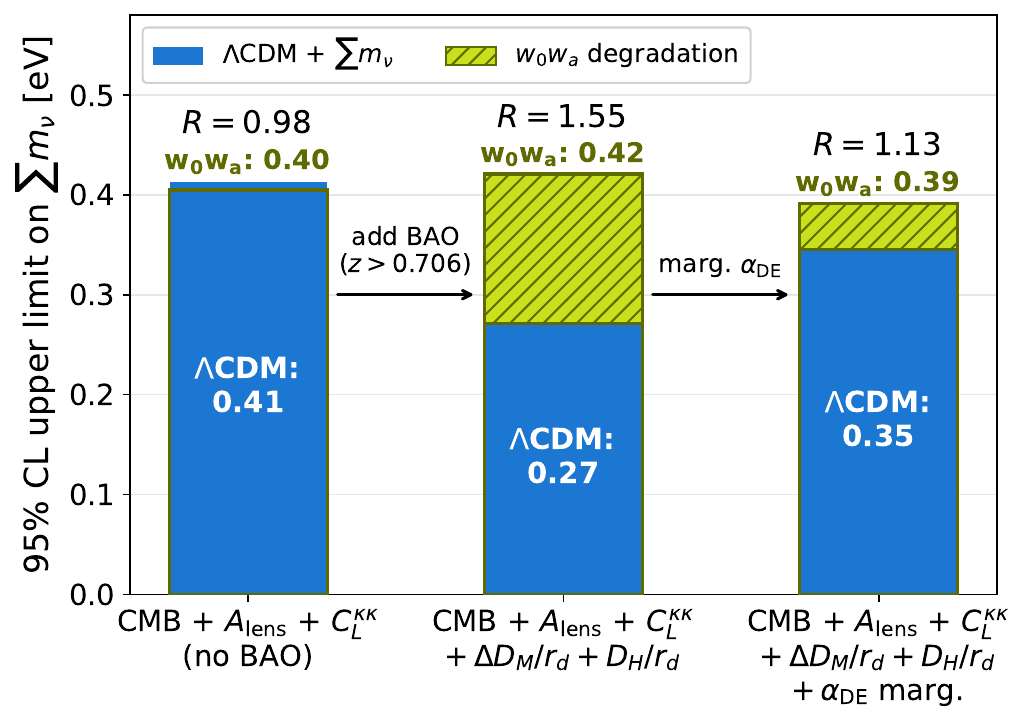}
\caption{Effect of adding radial BAO distances to the DE-insensitive
baseline (left bar: CMB + $A_{\rm lens}$ + $C_L^{\kappa\kappa}$ with
no BAO, identical to the left bar of Figure~\ref{fig:bao_bar}), and
the impact of marginalising over the late-time
distance parameter $\alpha_{\rm DE}$. Adding $\Delta D_M/r_d$ and
$D_H/r_d$ at $z > 0.706$ (middle bar) tightens the $\Lambda$CDM bound
from $0.41$ to $0.27$~eV but reintroduces dark-energy sensitivity
($R = 1.55$); since the differenced transverse distances carry no
information (Figure~\ref{fig:bao_bar}), the tightening is driven
entirely by $D_H/r_d$, and the constraint
is identical when only Ly$\alpha$ at $z = 2.33$ is retained.
Marginalising over $\alpha_{\rm DE}$ (right bar) restores most of the 
DE insensitivity ($R = 1.13$) and loosens the $\Lambda$CDM bound to 
$0.35$~eV. The mechanism is discussed in Section~\ref{sec:bao}. All 
constraints use mock data.}
\label{fig:dh_bar}
\end{figure}

The radial BAO distance $D_H(z)/r_d \equiv c/[H(z)\,r_d]$, combined
with the CMB-calibrated sound horizon, measures $H(z)$ at the tracer
redshift. In the matter-dominated regime
$H(z) \approx 100\sqrt{\omega_m(1+z)^3}$ is a one-parameter function
of the physical matter density $\omega_m = \Omega_m h^2$, so each
high-$z$ $D_H$ measurement is to leading order a measurement of
$\omega_m$. Adding $D_H/r_d$ at $z > z_{\rm ref}$ to the
$\Delta D_M/r_d$ baseline tightens the $\Lambda$CDM constraint from
$0.41$ to $0.27$~eV (Figure~\ref{fig:dh_bar}, middle bar); the
information saturates after one high-precision measurement and is
unchanged when only the Ly$\alpha$ point at $z = 2.33$ is retained.

The radial distances reintroduce dark-energy sensitivity despite measuring $H(z)$ in the matter-dominated regime: adding $D_H/r_d$ raises the ratio from $R = 1.02$ for the $\Delta D_M/r_d$ baseline to $R = 1.55$, and even the Ly$\alpha$ point at $z = 2.33$ alone gives $R = 1.54$. The sensitivity does not
originate in the radial distance itself: at $z = 2.33$ dark energy
contributes only ${\sim}6\%$ of $H^2$, so $D_H/r_d$ there measures
$H(z)$ deep in the matter-dominated regime and is nearly independent
of $w(a)$. It arises instead from the joint constraint that $D_H$ and
the CMB place on $\omega_m$, the channel through which late-time data
inform $\sum m_\nu$ (Appendix~\ref{sec:anatomy}): $D_H$ fixes $H(z)$ at
the tracer redshift, while the acoustic scale
$\theta_\star = r_\star/D_A(z_\star)$ ties $\omega_m$ to the
line-of-sight integral $\int_0^{z_\star} dz/H(z)$, whose low-redshift
portion is dark-energy-dominated. In $\Lambda$CDM, the two together
determine $\omega_m$; freeing $(w_0, w_a)$ loosens the low-redshift
integrand, $\theta_\star$ no longer constrains $\omega_m$, and
$\sigma(\omega_m)$ and $\sigma(\sum m_\nu)$ return to the no-$D_H$
baseline. The $\Lambda$CDM improvement from $0.41$ to $0.27$~eV is
therefore contingent on assuming a $\Lambda$CDM low-redshift expansion,
and the increase in $R$ quantifies the degradation once that assumption
is relaxed; this is the channel through which DESI's $D_H/r_d$
measurements drive the $w_0 w_a$ preference in the standard
combination.

Marginalizing over $\alpha_{\rm DE}$ confirms this by removing the
$\theta_\star$ constraint on the low-redshift integral directly. This
$\theta_\star$ constraint tightens the $\omega_m$ posterior in $\Lambda$CDM
but is degenerate with $(w_0, w_a)$ in $w_0 w_a$CDM; removing it therefore loosens only the
$\Lambda$CDM bound (from $0.27$ to $0.35$~eV), while the
$w_0 w_a$CDM bound stays at $\approx 0.39$~eV, bringing $R$ down to
$1.13$ (third bar of Figure~\ref{fig:dh_bar}). The residual $R > 1$
reflects $(w_0, w_a)$ freedom at the $D_H$ redshifts $z = 0.93$--$2.33$,
which lie above $z_{\rm DE} = 0.706$ and are therefore not absorbed by
$\alpha_{\rm DE}$.

The four-point lensing channel, by contrast, constrains the same
$\omega_m$ through the matter-era growth and geometry that set the
amplitude and shape of $C_L^{\kappa\kappa}$ (the horizon scale at
matter-radiation equality), with no dependence on a low-redshift
distance integral. Marginalizing $w(a)$ varies only the low-redshift
expansion, so it leaves the four-point constraint on $\omega_m$, and
hence on $\sum m_\nu$, unchanged.

In summary, every form of BAO information that improves the 
$\sum m_\nu$ constraint beyond the four-point lensing floor does so 
through geometric complementarity with the CMB-calibrated sound 
horizon, and this complementarity is inherently sensitive to the dark 
energy model. Whether the low-redshift information is removed from 
the BAO side (distance differencing), from the CMB side 
($\alpha_{\rm DE}$ marginalization, which broadens the $\theta_*$ 
posterior; Section~\ref{sec:unlensed}), or by freeing $(w_0, w_a)$ 
directly, the constraint returns to $\sim$0.41~eV with 
$R \approx 1$~\cite{Loverde_2024,Elbers:2025vlz}.

\section{Robustness to extended dark energy models beyond $\left( w_0, w_a\right)$}

Section~\ref{sec:lensing} established that the only dark-energy-robust neutrino-mass channels are the primary CMB and the four-point lensing reconstruction; the late-Universe-free combination of Section~\ref{sec:combined} is built from these alone. We now ask whether this combination is insensitive not merely to the two-parameter $(w_0, w_a)$ extension but to the broader class of smooth late-time expansion histories, testing it against progressively more flexible dark-energy models. 

\subsection{Dark energy-free neutrino masses: ideal mock data}
\label{sec:saturation}

\begin{figure}
\centering
\includegraphics[width=\columnwidth]{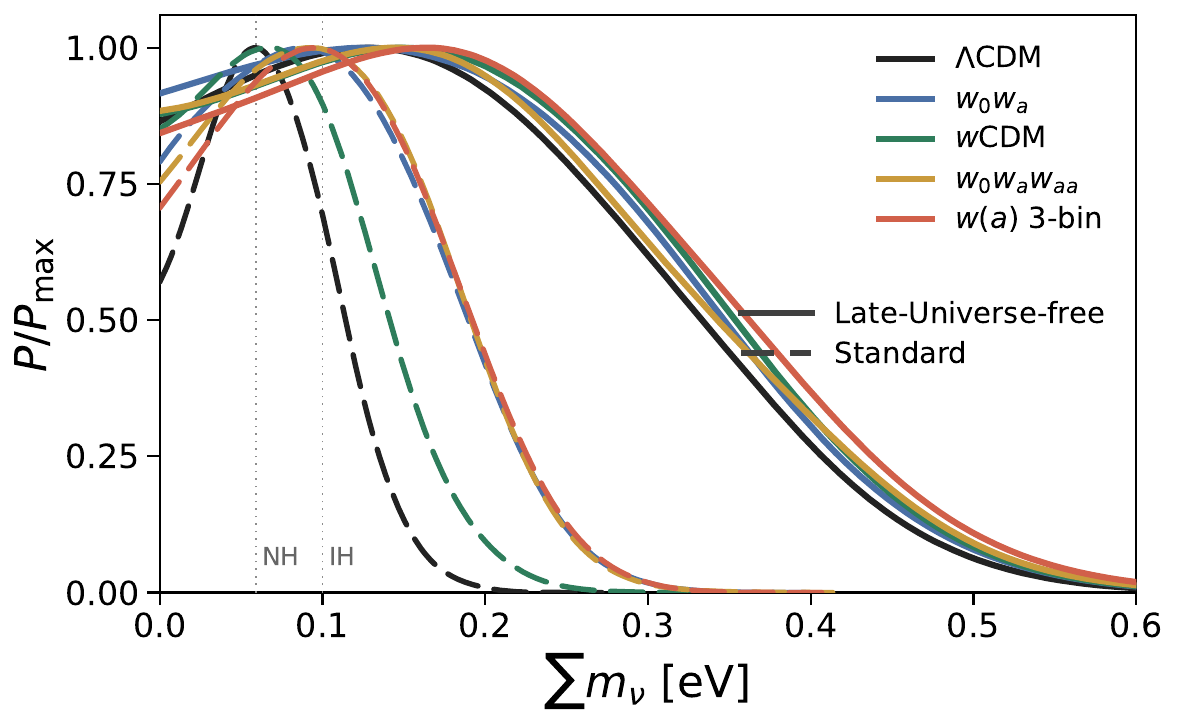}
\caption{$95\%$ upper limits on $\sum m_\nu$ on the mock
data (\textit{Planck} NPIPE CMB, CMB lensing, and DESI~DR2 BAO) across
dark-energy models of increasing flexibility, for the standard
combination (dashed) and the late-Universe-free combination (solid). Color
keys the dark-energy family, from $\Lambda$CDM through $w$CDM, $(w_0,
w_a)$, the cubic expansion $w_0 w_a w_{aa}$, and three-bin $w(a)$. The
standard bound degrades from $\Lambda$CDM to $(w_0, w_a)$ and then
saturates; the late-Universe-free bound is flat across all models by
construction. Vertical dotted lines mark the normal- (NH) and
inverted-ordering (IH) floors.}
\label{fig:saturation_fiducial}
\end{figure}

Working on the mock data, whose underlying truth is the $\Lambda$CDM fiducial cosmology of Section~\ref{sec:metric} with $\sum m_\nu = 0.06$~eV, we vary the background expansion across a sequence of increasing freedom: a constant equation of state $w$CDM; the CPL form $w(a) = w_0 + w_a(1-a)$; piecewise-constant $w(a)$ in three and five redshift bins, which relaxes the smooth-evolution assumption of CPL; a cubic expansion of $w(a)$. For each we track the $95\%$ upper limit on $\sum m_\nu$ (Figure~\ref{fig:saturation_fiducial}).

The late-Universe-free bound is flat across the entire sequence, holding at $0.40$--$0.42$~eV. This flatness is not fully guaranteed by construction: the combination retains no distance information for the parametrization to act on, but dark energy still enters $C_L^{\kappa\kappa}$ through the low-redshift growth factor and the lensing-kernel distances. The flat bound demonstrates that this residual growth-level channel is negligible, consistent with the per-probe result of Section~\ref{sec:lensing_de_sensitivity}. Since the mock data are generated within $\Lambda$CDM and every extended family nests it, the best-fit $\chi^2$ is unchanged by construction ($\Delta\chi^2 = -0.4$ and $+0.1$ for $(w_0, w_a)$ and cubic $w(a)$ relative to $\Lambda$CDM, consistent with noise in the minimization); the mock exercise therefore isolates the response of the bound to parameter degeneracies and marginalization volume alone, uncontaminated by any dark-energy pull in the data. Whether the real data, which do carry such a pull, move the bound is the subject of Section~\ref{sec:combined}.

\begin{figure}[t]
\centering
\includegraphics[width=\columnwidth]{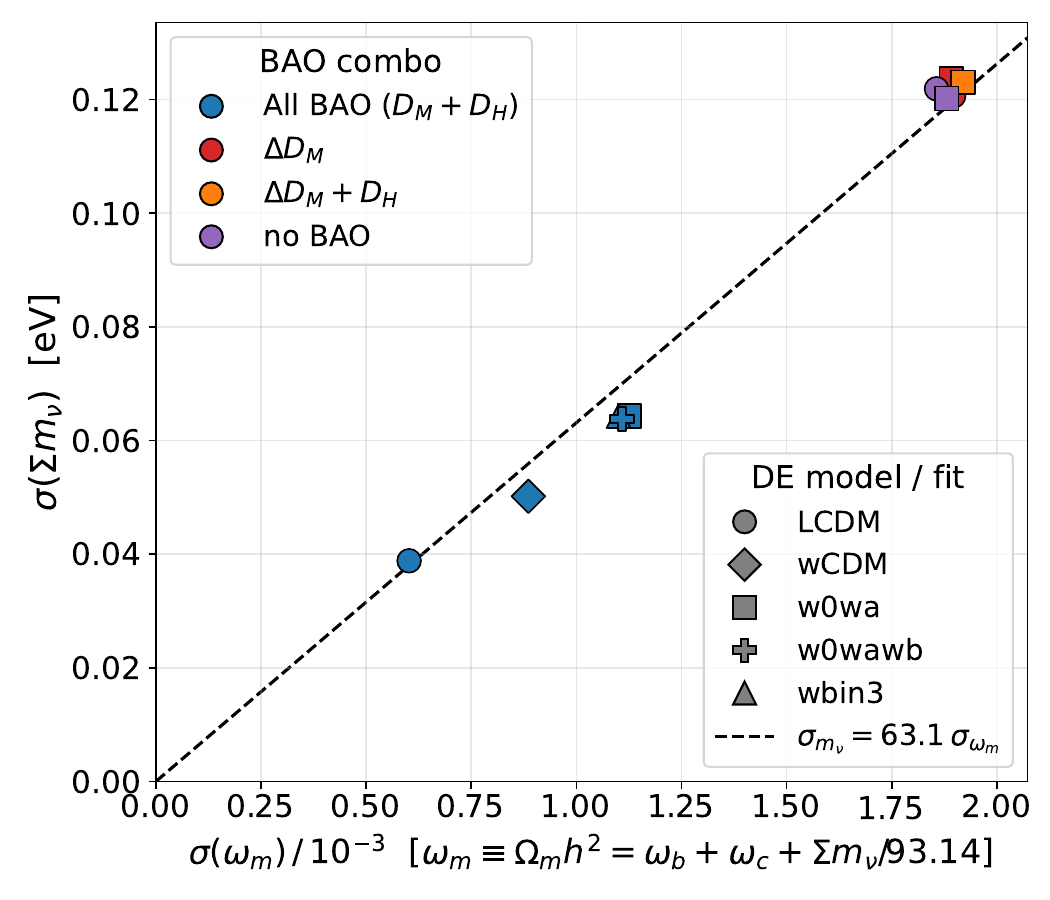}
\caption{
Marginal $\sigma(\sum m_\nu)$ versus $\sigma(\omega_m)$ for ten mock-data chains spanning a threefold range in precision and covering $\Lambda$CDM, $w$CDM, $(w_0,w_a)$, and binned and cubic $w(a)$ with all-BAO ($D_M+D_H$), differenced ($\Delta D_M$ and $\Delta D_M+D_H$), and no-BAO distance treatments; $A_{\rm lens}$ is fixed at unity in the all-BAO (standard) combination and marginalized in the others, and all chains include the four-point reconstruction $C_L^{\kappa\kappa}$. The points fall on a single near-linear locus through the origin with slope $k \approx 63.1$~eV (Eq.~\ref{eq:lockstep}, Appendix~\ref{sec:anatomy}), so the neutrino-mass information carried by late-time data is set by how well they pin the matter density.}
\label{fig:omegam_lockstep}
\end{figure}

We now turn to the standard combination, which retains the full distance information rather than discarding it.
Here the question is whether marginalizing
over the parameters $w_0, w_a$
suffices to absorb the dark-energy sensitivity, or whether, as the standard objection holds, a more flexible expansion history would keep weakening the bound. We find the former. The bound degrades sharply from $\Lambda$CDM to $(w_0, w_a)$, from $0.135$ to $0.219$~eV, but then saturates rather than continuing to weaken: three-bin $w(a)$, five-bin $w(a)$, and cubic CPL give $0.221$, $0.207$, and $0.217$~eV, all within $6\%$ of the $(w_0, w_a)$ value. This behaviour inherits the matter-density saturation of Appendix~\ref{sec:anatomy}: late-time data inform $\sum m_\nu$ through the physical matter density, the neutrino contribution entering as the residual between the total and the CMB-calibrated cold dark matter and baryon densities, $\sum m_\nu \propto \omega_m - \omega_{cb}$~\cite{Loverde:2024nfi}, so the $\sum m_\nu$ bound tracks $\sigma(\omega_m)$ (Figure~\ref{fig:omegam_lockstep}). That uncertainty rises from $\sigma(\omega_m) = 6\times10^{-4}$ in $\Lambda$CDM to $1.1\times10^{-3}$ under $(w_0, w_a)$, compared with $1.9\times10^{-3}$ when the BAO distances are removed entirely, and remains at $1.1\times10^{-3}$ under binned and cubic $w(a)$, since the additional $w(a)$ modes are nearly orthogonal to the $\omega_m$ direction constrained by the CMB.

This plateau is not an artifact of the additional dark-energy parameters being unconstrained. Decomposing the posterior of each model into the eigenmodes of its $w(a)$ parameters, normalized to the prior widths, the data resolve three to four independent modes: the three-bin model has eigenmode ratios $\sigma_{\rm post}/\sigma_{\rm prior} = 0.11, 0.36, 0.53$, all three data-constrained, and the five-bin model $0.12, 0.36, 0.46, 0.69, 1.02$, three to four constrained. The data therefore measure more dark-energy freedom than $(w_0, w_a)$ provides; the additional resolved modes are simply nearly orthogonal to the CMB-bounded $\omega_m$ anchor (Appendix~\ref{sec:anatomy}) and carry essentially no further neutrino-mass information. An independent singular-value decomposition of the DESI DR2 BAO and Type Ia supernova distances reaches the same conclusion from a purely geometric standpoint: the only direction these data measure beyond the CMB is $\omega_m$, and the $w_0 w_a$ preference is dominated by this single mode, with the orthogonal dark-energy direction consistent with a cosmological constant~\citep{zaldarriaga2026universaldistancemodesdesi}. Within smooth late-time expansion histories, $(w_0, w_a)$ marginalization is therefore sufficient: it already spans the dark-energy directions to which $\sum m_\nu$ is sensitive, and more heavily parametrized backgrounds produce no further degradation.

 While this stability has been noted empirically, in a binned-$w(z)$ analysis of DESI~DR1 data whose neutrino-mass bounds were no weaker with five or ten redshift bins opened, an effect left unexplained and tentatively ascribed to the positivity prior acting in the enlarged parameter space~\citep{reboucas2025desiy1neutrinos}, in node-based and transition-feature reconstructions whose bounds fall at or below the $(w_0, w_a)$ value without a direct comparison~\citep{ghedini2025wreconstruction, nair2025fourparamde, barua2025secondordercpl}, and in a full-shape reanalysis of DESI~DR1 which bounds $\sum m_\nu$ under $(w_0, w_a)$ while noting evidence that higher-order terms in the $w(a)$ expansion do not improve the fit~\citep{chudaykin2026reanalyzingdesidr12}, a controlled demonstration has to our knowledge been missing: a nested hierarchy of $w(a)$ parametrizations analysed on a common dataset and pipeline, with the additional expansion modes shown to be resolved by the data and the stability traced to its origin in the $\omega_m$ anchor; the analysis above provides this demonstration.

Both results are statements about smooth, matter-conserving
dark energy that modifies only the late-time expansion. For this class
the saturation establishes $(w_0, w_a)$ marginalization as a sufficient
treatment, and the late-Universe-free bound of Section~\ref{sec:combined}
limits the exposure further by construction: having retained no absolute
distance information, it offers no $\omega_m$ channel through which a
misspecified expansion history could displace the inferred mass. Neither
statement covers dark-energy physics outside this class: a dark sector
that exchanges energy with matter shifts $\omega_m$ itself; modifications
that alter the pre-recombination sound horizon $r_d$ bias the absolute
and differenced BAO alike; and dark energy that clusters (effective sound
speed $c_s^2 \ll 1$) is degenerate with neutrino free-streaming at the
perturbation level, displacing the inferred mass even at fixed background
expansion, in current data toward smaller and even negative effective
masses, tightening the nominal limit rather than relaxing
it~\cite{yang2026darkenergyperturbationsrobustness}. We likewise take the
neutrino sector to be standard, three degenerate active states with
$N_{\rm eff} = 3.044$ interacting only gravitationally; extended neutrino
sectors are beyond the scope of this test.

\subsection{Dark energy-free neutrino mass: real data}
\label{sec:combined}

The preceding sections, carried out on mock data to avoid noise
fluctuations, anomalies, or residual systematics, identified a single
combination that carries meaningful neutrino-mass information while
remaining insensitive to the dark-energy model: the primary CMB with
$A_{\rm lens}$ marginalization, combined with the reconstructed
four-point lensing power spectrum $C_L^{\kappa\kappa}$. BAO is excluded
not because it carries no information on $\sum m_\nu$, but because that
information is inseparable from dark-energy sensitivity: it enters
through the $\omega_m$ anchor set by comparing distances across the
matter- and dark-energy-dominated eras, which dark-energy freedom
degrades, and its dark-energy-insensitive residue, the differenced
$\Delta D_M/r_d$, adds nothing beyond the CMB and lensing determination
of $\omega_m$ (Section~\ref{sec:bao}). We now apply this combination to
the real data and set it against the standard analysis.

The standard combination of Planck NPIPE CMB, ACT DR6 + SPT + Planck
lensing, and DESI DR2 BAO yields
\begin{equation}
    \sum m_\nu < 0.056 \ \mathrm{eV \ (95\% \ C.L.,\ CMB + CMBL + BAO)},
\end{equation}
in $\Lambda$CDM, in $\sim 2$--$3\sigma$ tension with the inverted-ordering floor ($\sum m_\nu \gtrsim 0.10$~eV). This bound depends on the assumed
dark-energy background and weakens as more late-time expansion freedom
is allowed: one equation-of-state parameter ($w$CDM) relaxes it to
$\sum m_\nu < 0.080$~eV and the full $(w_0, w_a)$ extension to
$\sum m_\nu < 0.152$~eV, inflating $\sigma(\sum m_\nu)$ by $R_w = 1.4$
and $R_{w_0 w_a} = 2.5$ respectively. The standard limit therefore spans
a factor of $2.5$ in $\sigma(\sum m_\nu)$, fixed by nothing more than the
chosen dark-energy parametrization.

Applying instead the late-Universe-free combination identified above,
we obtain
\begin{equation}
\begin{split}
    \sum m_\nu < 0.41 \ &\mathrm{eV \ (95\% \ C.L.,} \\
    &\mathrm{CMB + A_{\rm lens} + CMBL + \Delta D_M/r_d)},
\end{split}
\end{equation}
in $\Lambda$CDM, with the bound shifting by less than $4\%$ across all
three dark-energy models ($0.41$, $0.41$, $0.40$~eV for $\Lambda$CDM,
$w$CDM, and $w_0 w_a$CDM; the underlying $\sigma(\sum m_\nu)$ agree to
$2\%$, $R_w = 0.99$, $R_{w_0 w_a} = 0.98$). 

\begin{figure}
\centering
\includegraphics[width=\columnwidth]{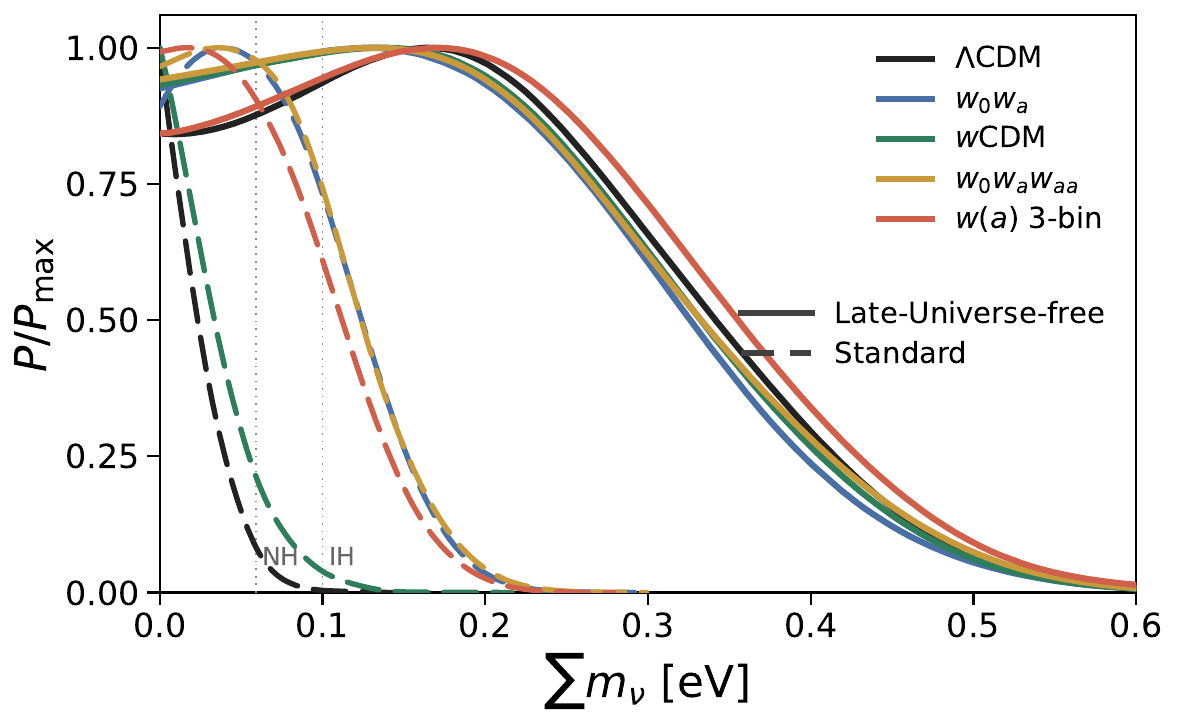}
\caption{$95\%$ upper limits on $\sum m_\nu$ on the real data
(\textit{Planck} NPIPE CMB, CMB lensing, and DESI~DR2 BAO) across
dark-energy models of increasing flexibility, for the standard
combination (dashed) and the late-Universe-free combination (solid;
CMB $+\,A_{\rm lens}+C_L^{\kappa\kappa}+\Delta D_M/r_d$). Color keys
the dark-energy family, from $\Lambda$CDM through $w$CDM, $(w_0, w_a)$,
the cubic expansion $w_0 w_a w_{aa}$, and three-bin $w(a)$. The standard
combination yields $\sum m_\nu < 0.056$~eV in $\Lambda$CDM, in tension
with the oscillation floor, broadening to $< 0.080$~eV under $w$CDM and
to $< 0.152$~eV under $(w_0, w_a)$, then saturating ($< 0.155$ and
$< 0.146$~eV under $w_0 w_a w_{aa}$ and three-bin $w(a)$). The
late-Universe-free bound is flat across all models, holding at
$0.40$--$0.43$~eV, demonstrating that it is independent of the assumed
late-time dark-energy parametrization. Vertical dotted lines mark the
normal- (NH) and inverted-ordering (IH) floors.}
\label{fig:mnu_1d_data}
\end{figure}

This bound is insensitive to these choices of dark energy models.
This bound is $\sim 7$ times larger than the $\Lambda$CDM bound of $0.056$~eV.
However, its cost is most fairly stated against the standard bound granted the
same dark-energy robustness: relative to the $(w_0, w_a)$-marginalized
$0.152$~eV, the late-Universe-free limit is weaker by a factor of $2.7$ on
the real data ($1.9$ on mock data). 
Figure~\ref{fig:mnu_1d_data} shows the contrast: the
standard posteriors fan out from $\Lambda$CDM through $w$CDM to
$(w_0, w_a)$ as the dark-energy freedom opens up, while the
late-Universe-free posteriors for all three models lie nearly on top of
one another.

The late-Universe-free bound is also stable between mock and real data ($0.405$ and $0.41$~eV), whereas the standard bound is not: in
$\Lambda$CDM it moves from $0.135$ to $0.056$~eV, and under $(w_0, w_a)$ from $0.219$ to $0.152$~eV, tightened on the real data by the tension between the CMB and BAO distances. 
Part of the standard limit's strength rests on a tension whose interpretation is itself in question (Section~\ref{sec:metric}).

The same data express a preference for dark energy different from $\Lambda$ in the standard combination~\citep{desicollaboration2025desidr2resultsii,garciaquintero2025cosmologicalimplicationsdesidr2}, and this preference, like the bound, is
exhausted at $(w_0, w_a)$. Relative to $\Lambda$CDM, the best-fit
$\chi^2$ improves by $5.5$ under $(w_0, w_a)$, with the gain concentrated
in the BAO component (from $12.0$ to $7.2$); more flexible models do not
improve the fit further ($\Delta\chi^2 = -5.9$ for three-bin and $-5.6$
for cubic $w(a)$ relative to $\Lambda$CDM, essentially unchanged from
$(w_0, w_a)$). This is the same low-redshift distance preference that
anchors $\omega_m$~\citep{Loverde:2024nfi,Lynch:2025ine}, and it is what drives the standard neutrino bound down. The
late-Universe-free combination of Section~\ref{sec:combined} expresses no
such preference: its best-fit $\chi^2$ varies by $\lesssim 2$ across all
dark-energy models, with no systematic improvement. Having removed the dark-energy-sensitive distance information by construction, it retains
nothing with which to prefer one dark-energy model over another.

The two routes also differ in what they assume about the measurements
themselves. The $(w_0, w_a)$ marginalization retains the absolute
low-redshift distances and attributes any departure from $\Lambda$CDM
expansion entirely to dark energy; an unmodeled systematic in the
distance--redshift relation would therefore be absorbed into
$(w_0, w_a)$ and propagated into $\sum m_\nu$. The late-Universe-free
construction discards this information by construction, and is thus
insensitive not only to the dark-energy model but to any coherent
distortion of the low-redshift distance scale, at the price of a bound
weaker by a factor of $2.7$. At present this caution is not driven by
any anomaly: the DESI DR2 measurements pass their internal consistency
tests~\citep{desicollaboration2025desidr2resultsii}. Going forward,
as the standard bound tightens toward the oscillation floors, the
comparison with the distance-free bound becomes a natural consistency
check.

\section{Future outlook: Rubin \& Simons observatories, Spec-S5 and beyond}
\label{sec:forecast}

\subsection{Simons Observatory \& Spec-S5: the late-Universe-free combination remains robust and improves}

\begin{figure}
\centering
\includegraphics[width=\columnwidth]{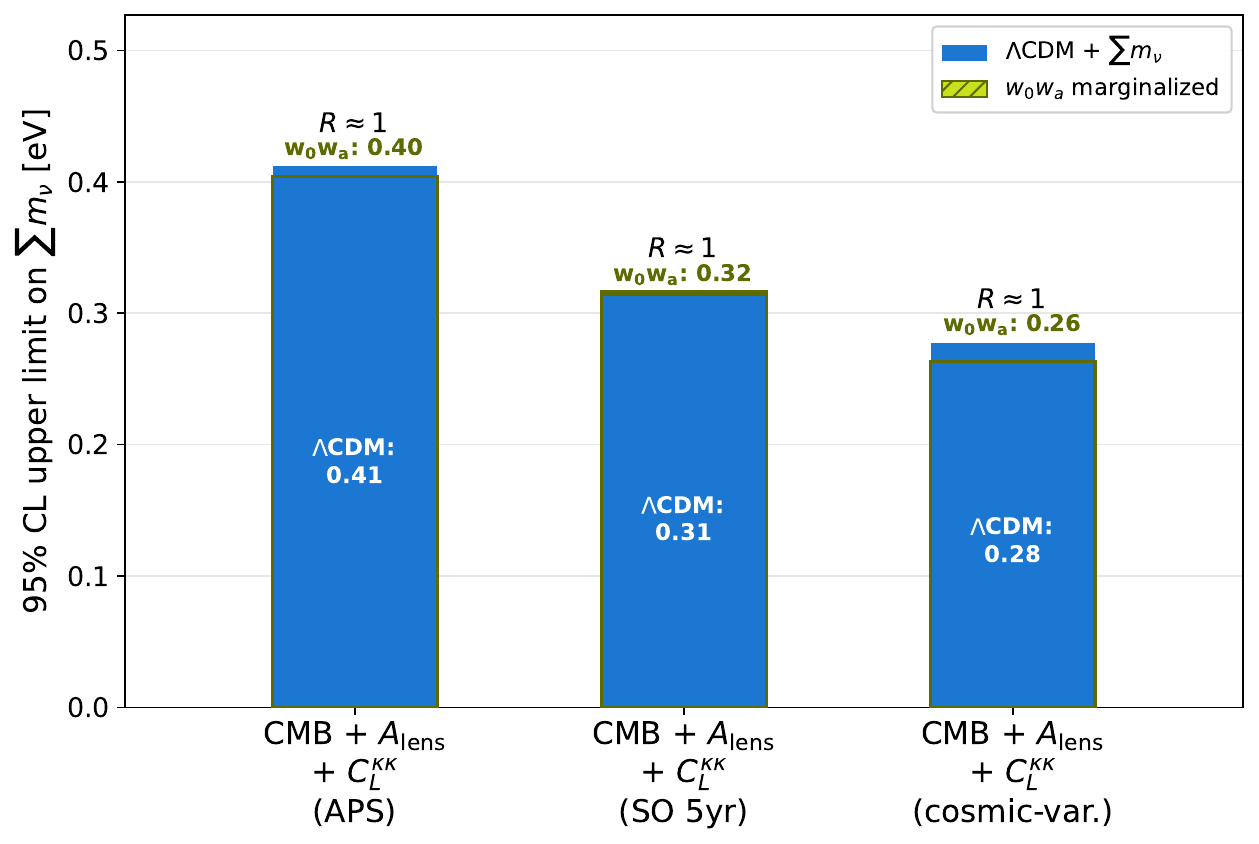}
\caption{Forecast $95\%$ CL upper limits on $\sum m_\nu$ for the
late-Universe-free channel (NPIPE primary CMB with $A_{\rm lens}$ marginalized,
plus the four-point CMB lensing reconstruction alone; mock data
throughout, generated at $\sum m_\nu = 0.06$~eV), as the lensing sensitivity advances from the current
ACT+\textit{Planck}+SPT reconstruction (APS) through a five-year Simons
Observatory-like reconstruction to the cosmic-variance limit.
Marginalizing over $(w_0, w_a)$ (green hatched) leaves the bound
unchanged at every noise level ($R \approx 1$ denotes
$|R - 1| \leq 0.06$, consistent with unity within chain sampling
noise); blue bars are the $\Lambda$CDM limit and the
$(w_0, w_a)$-marginalized limit is annotated above each bar. For
reference, the standard combination (lensed CMB, four-point CMB lensing,
and BAO) advancing over the same period from current data
(\textit{Planck} + DESI DR2) to forecast precision (five-year Simons
Observatory-like lensing + Spec-S5 BAO) is degraded by $R \simeq 1.5$--$1.6$
under the same $(w_0, w_a)$ marginalization; its bound is discussed in
the text. The corresponding marginalized posterior widths
for the late-Universe-free bars are $\sigma(\sum m_\nu) = 0.12$, $0.093$,
and $0.082$~eV; on these mock data the plotted limits sit at
$\approx 3.4\,\sigma(\sum m_\nu)$ above zero (Section~\ref{sec:metric}),
and comparisons with Fisher forecasts should be made in $\sigma$.}
\label{fig:forecast_grouped}
\end{figure}

In the preceding sections we identified the unlensed primary CMB and the four-point CMB lensing as the
only  dark-energy-insensitive channel carrying meaningful
$\sum m_\nu$ information. We now ask how tight a dark-energy-agnostic
bound this channel delivers at the substantially higher signal-to-noise
of next-generation experiments. We forecast the combination holding the
primary CMB at its current precision and replacing the BAO and
CMB-lensing likelihoods with mock data at the projected precision
of a Spec-S5-like survey ($0.18$--$1\%$ on the BAO distance scale
across $0 < z < 5$) and a five-year Simons Observatory-like minimum-variance
lensing reconstruction \citep{2019JCAP...02..056A} (total signal-to-noise $\sim 160$, roughly twice
the present ACT+\textit{Planck}+SPT value).

The dark-energy-agnostic combination remains robust across dark-energy
models (Figure~\ref{fig:forecast_grouped}; $R = 1.01$;
$\sigma(\sum m_\nu) = 93$~meV; $\sum m_\nu < 0.314$~eV in $\Lambda$CDM, $0.316$~eV under
$(w_0, w_a)$, and $0.321$~eV under cubic $w(a)$), tightening from its
present $\sigma(\sum m_\nu) = 0.12$~eV ($\sum m_\nu < 0.41$~eV) through the improved lensing reconstruction; its
insensitivity to the late-time expansion therefore survives the gain
in lensing sensitivity. Pushing the four-point reconstruction to the cosmic-variance limit
tightens the bound only modestly further, to $\sigma(\sum m_\nu) = 82$~meV, $\sum m_\nu < 0.28$~eV
($R \approx 1$); the five-year Simons Observatory reconstruction
($0.31$~eV) therefore already captures most of the four-point
neutrino-mass information, with the residual gap to the
cosmic-variance limit small and the dark-energy insensitivity retained
at both noise levels.

For the standard combination, the $95\%$
upper limit relaxes from $\sum m_\nu < 0.100$~eV ($\sigma = 27$~meV) in $\Lambda$CDM to
$0.153$~eV ($\sigma = 43$~meV) once $(w_0, w_a)$ are marginalized; $(w_0, w_a)$ saturates
the bound, and even at this precision more extended models do not
degrade it further; the neutrino-dark-energy degeneracy is spanned by
$(w_0, w_a)$ and does not widen as precision improves.

The forecast standard-combination bound of $0.153$~eV improves on its mock-current counterpart ($0.219$~eV) by $30\%$ but only matches the
current real-data bound ($0.152$~eV), which is tightened by the same low-redshift distance pull that drives the dark-energy preference and is absent from the mock; the late-Universe-free bound is instead stable between mock and real data ($0.41$~eV on both), so its improvement to
$0.31$~eV reflects the gain in lensing signal-to-noise alone. At this precision the standard combination reaches $\sigma(\sum m_\nu) = 27$~meV within $\Lambda$CDM, placing the minimal
$60$~meV mass sum at $2.2\sigma$, but degrades to $43$~meV under $(w_0, w_a)$ marginalization and to $93$~meV for the late-Universe-free
combination; a resolution of the present preference for sub-minimal $\sum m_\nu$ therefore remains conditional on the assumed late-time expansion history.

\subsection{Combining Rubin \& Simons observatories}

A further question is whether other large-scale structure probes can supply dark-energy-agnostic $\sum m_\nu$ information beyond the
four-point channel. Galaxy weak lensing tomography, galaxy clustering, and their cross-correlations with each other and with CMB lensing probe both growth and geometry across a broad redshift range \citep{Mishra_Sharma_2018}, with sensitivity to $\sum m_\nu$ and to dark energy that differs from BAO. The
constraining power of these combinations is already demonstrated:
\citep{PhysRevD.111.083516} obtain $\sum m_\nu < 0.12$~eV from a joint
analysis of ACT~DR6 + \textit{Planck}~PR4 lensing, unWISE galaxy
clustering, BAO, and primary CMB.

To assess the dark-energy degradation at future precision we
perform a Fisher forecast for three LSST + Simons Observatory  dataset
combinations following exactly \citep{Schaan_2020}, each built on a Planck
CMB prior alone with no BAO information added, in keeping with the
late-Universe-free framing of the previous sections: the LSST $3\times 2$-point
with the cosmic shear sector ($\gamma\gamma, \gamma g, gg$); an
alternative $3\times 2$-point in which the shear pairs are replaced
by CMB lensing cross-correlations ($\kappa\kappa, \kappa g, gg$);
and the full $6\times 2$-point analysis combining all six spectra.
The Planck CMB prior is constructed from the covariance of the same
Planck-only MCMC chain ($\Lambda$CDM + $\sum m_\nu + w_0 + w_a +
A_{\rm lens}$) used elsewhere in this work. We assume the LSST Y10
Gold sample \citep{lsstsciencecollaboration2009lsstsciencebookversion}
(10 tomographic bins, $f_{\rm sky} = 0.35$, $k_{\rm max} = 0.3\,h/$Mpc,
$\ell_{\rm max} = 1000$) and a five-year Simons Observatory-like
minimum-variance lensing reconstruction
\citep{2019JCAP...02..056A}, and marginalize over 130
nuisance parameters.

\begin{figure}
\centering
\includegraphics[width=\columnwidth]{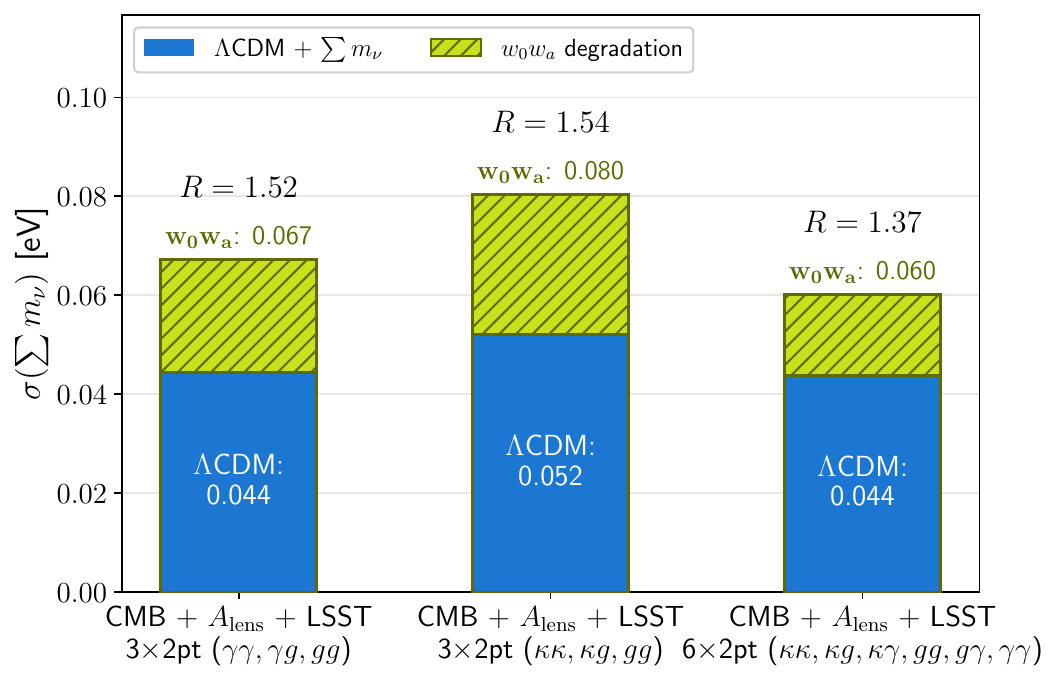}
\caption{Fisher forecast for the degradation of the neutrino mass constraint
when dark energy parameters ($w_0$, $w_a$) are freed, for three
dataset combinations built on a Planck CMB prior with $A_{\rm lens}$
marginalized; no BAO information is included. All three combinations
adopt the same survey configuration: 10 tomographic bins over
14,300~deg$^2$, marginalizing over 130 nuisance parameters (galaxy
bias, shear calibration, and Gaussian + outlier photo-$z$). The first
bar adds the LSST $3\times 2$-point analysis with the cosmic shear
sector ($\gamma\gamma$, $\gamma g$, $gg$); the second replaces the
shear-based pairs with Simons Observatory lensing cross-correlations to form an
alternative $3\times 2$-point ($\kappa\kappa$, $\kappa g$, $gg$); the
third combines all six spectra ($\kappa\kappa$, $\kappa g$,
$\kappa\gamma$, $gg$, $g\gamma$, $\gamma\gamma$) into the full
$6\times 2$-point analysis. Blue bars show the Fisher width
$\sigma(\sum m_\nu)$ in $\Lambda$CDM; green hatched regions show the
additional degradation when $w_0$ and $w_a$ are varied (see
Section~\ref{sec:metric} for the relation to reported credible
limits). The ratio $R$ quantifies
the degradation factor. At Simons Observatory lensing noise the cosmic-shear
and $\kappa$-cross $3\times 2$-point combinations degrade comparably
($R = 1.52$ and $1.54$), while combining all six spectra reduces the
degradation to $R = 1.37$ and tightens the $\Lambda$CDM bound. The
forecast setup follows  \citet{Schaan_2020}.}
\label{fig:forecast}
\end{figure}

The results are shown in Figure~\ref{fig:forecast}. In $\Lambda$CDM
all three combinations yield $\sigma(\sum m_\nu) \approx 45$--$52$~meV%
; in the credible-limit convention of
Figure~\ref{fig:forecast_grouped} these widths correspond to reported
limits of $\approx 0.15$--$0.18$~eV on a minimal-mass sky
(Section~\ref{sec:metric}).

These forecast widths are comparable to the
$\sigma(\sum m_\nu) = 39$~meV ($0.135$~eV credible limit) obtained from
the standard current combination on mock $\Lambda$CDM data, and the
corresponding reported limits sit well above the current real-data
bound $\sum m_\nu < 0.056$~eV:
the real-data limit is tighter only because the data prefer an
effective neutrino mass below zero,
$\sum m_{\nu,{\rm eff}} \simeq -0.10$~eV~\citep{Elbers:2025vlz}, which the
positivity prior converts into an anomalously tight upper bound. The
value of the future measurement is therefore not raw precision but
independence: the $6\times2$-point is independent of BAO, and
analyzed with the positivity prior relaxed, it would measure the
effective mass to $\pm 45$--$52$~meV. The separation between the
current preference ($-0.10$~eV) and the minimal normal-hierarchy mass
($+0.06$~eV) is $0.16$~eV, so this combination would discriminate the
two at the $\sim 3\sigma$ level, directly testing whether the current
tight bounds reflect the neutrino mass or a low-redshift anomaly.
Freeing $w_0$ and $w_a$ degrades the cosmic-shear $3\times 2$-point by
$R = 1.52$; at Simons Observatory lensing noise the $\kappa$-cross
$3\times 2$-point degrades comparably ($R = 1.54$), and the full
$6\times 2$-point reaches the lowest degradation, $R = 1.37$, with
$\sigma(\sum m_\nu) = 60$~meV (a reported limit of
$\approx 0.20$~eV in the credible-limit convention). Cosmic shear is sourced primarily at
$z \lesssim 1$ where dark energy dominates, whereas CMB lensing is
sourced at $z \sim 2$ where dark energy is negligible; combinations
including $\kappa$-cross spectra inherit the high-redshift weighting,
in line with the four-point result of Section~\ref{sec:lensing}, and
the advantage of the $6\times 2$-point over the shear-only
$3\times 2$-point reflects this weighting.
For comparison, the standard CMB + lensing + BAO combination
degrades by $R_{w_0 w_a} = 2.5$; the LSST + Simons Observatory
$6\times 2$-point reduces this to $1.37$ without using BAO. 

\begin{figure}
\centering
\includegraphics[width=\columnwidth]{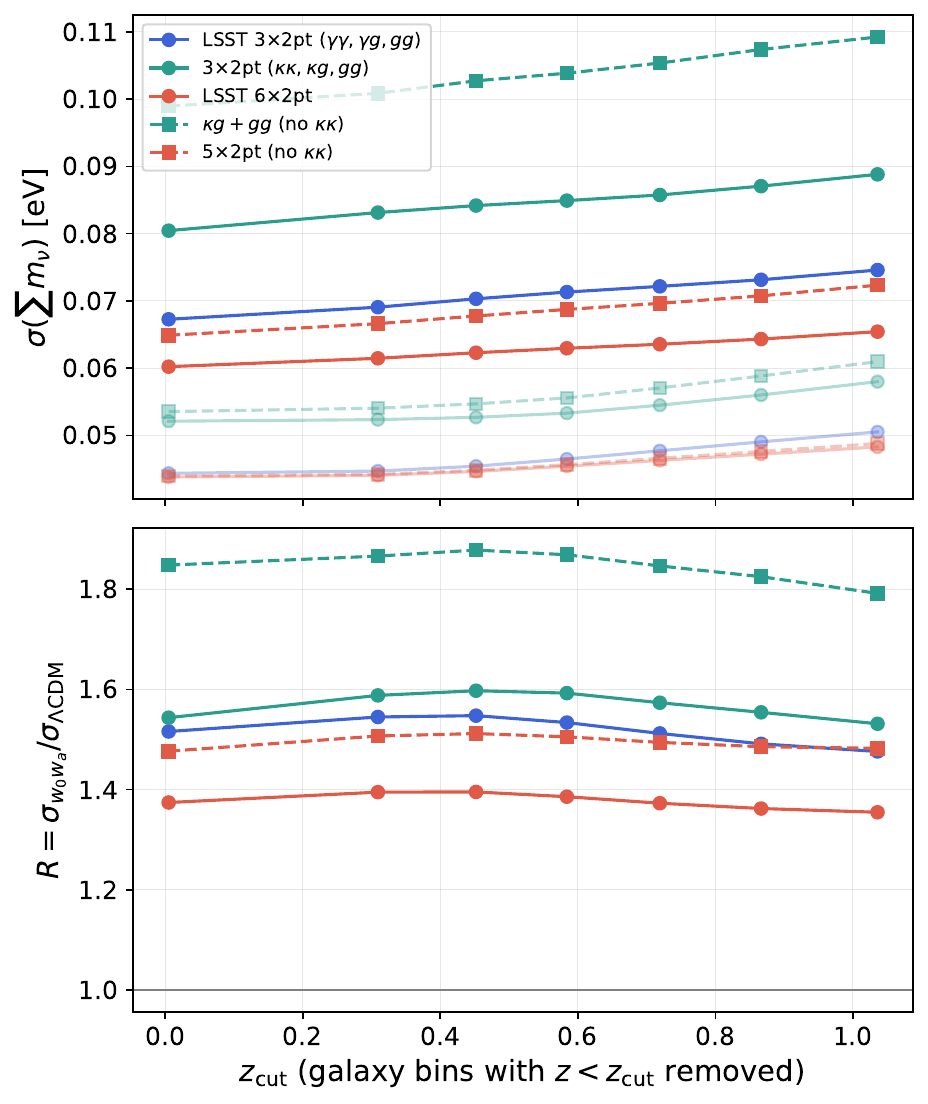}
\caption{Effect of removing low-redshift galaxy bins on the dark-energy
degradation. The LSST clustering bins with $z < z_{\rm cut}$ are dropped
from each combination and the forecast is recomputed at every cut.
\emph{Top:} the Fisher width $\sigma(\sum m_\nu)$ for
$\Lambda$CDM (faded) and $w_0 w_a$CDM (opaque); both widths increase as
bins are removed, but their ratio remains nearly constant. \emph{Bottom:} the degradation factor
$R = \sigma_{w_0 w_a}/\sigma_{\Lambda{\rm CDM}}$, which is flat to mildly
falling rather than dropping toward unity. The low-redshift galaxy bins
self-calibrate $(w_0, w_a)$, so discarding them de-constrains dark energy
and increases the marginalization cost rather than removing it; the
dark-energy sensitivity is a property of the combination, not a separable
low-$z$ data portion. The full $6\times 2$-point reaches the lowest $R$,
and dropping the $\kappa\kappa$ auto-spectrum raises $R$ sharply (dashed
curves) regardless of the low-$z$ cut, so it is the high-redshift
$\kappa$ information, not the removal of low-$z$ data, that limits the degradation.}
\label{fig:lowz_cut}
\end{figure}

A simpler attempt to remove the dark-energy sensitivity is to discard
the low-redshift galaxy bins outright, since dark energy acts only at
$z \lesssim z_{\rm DE}$.\footnote{Although not applied here, the
analogous cleaning can be performed on the CMB lensing convergence
itself: the $z \lesssim 1$ contribution to the reconstructed $\kappa$
can be nulled at the field level by subtracting suitably weighted
galaxy density tracers \citep{Qu_2022, Baleato_Lizancos_2023}.} We
test this by progressively removing the LSST clustering bins below a
cut $z_{\rm cut}$ from the $3\times 2$-point and $6\times 2$-point
combinations and recomputing $R$ at each step
(Figure~\ref{fig:lowz_cut}). The degradation does not shrink: $R$ is
almost flat as low-redshift bins are dropped. The mechanism is the
same self-calibration identified for BAO differencing
(Section~\ref{sec:bao}): the low-redshift galaxy bins carry the
internal constraint on $(w_0, w_a)$, so removing them broadens the
dark-energy posterior and increases the volume over which the
marginalization is performed, while the $\Lambda$CDM bound is nearly
unchanged because the neutrino-mass information is not localized at
low redshift; it is set instead by the $\omega_m$ anchor and the
broad lensing kernels. The dark-energy sensitivity of these probes is therefore a property of the combination, not of a separable
low-redshift data portion that can be removed; it is the
high-redshift weighting of the $\kappa$-cross spectra, rather than a
low-$z$ cut, that limits $R$.

\subsection{Benefits of cleaning low-z CMB lensing with galaxies}

Our approach connects naturally to \citep{Qu_2022}, where it was
proposed to null the low-redshift contribution to the reconstructed
CMB lensing convergence with galaxy tracers; in their configuration
our pipeline reproduces their forecast uncertainty
($\sigma(\sum m_\nu) = 35$~meV against their published $39$~meV). Testing the proposal as an explicitly
dark-energy-marginalized constraint on mock $\Lambda$CDM data, with
Planck-precision CMB ($A_{\rm lens}$ marginalized), the $z > 5$
contribution to $C_L^{\kappa\kappa}$ at their CMB-S4 noise ($L \le 200$),
and a single $z = 4.5$ absolute BAO pair at MegaMapper precision, we
obtain $\sum m_\nu < 0.36$~eV in $\Lambda$CDM, degrading only to
$0.39$~eV with $(w_0, w_a)$ freed: $R = 1.07$. The two high-redshift
ingredients ($z > 5$ lensing and $z = 4.5$ BAO) are required jointly:
 replacing either with its
low-redshift-contaminated counterpart breaks the insensitivity.
Substituting the uncleaned $C_L^{\kappa\kappa}$ for its $z>5$
contribution, with the same $z=4.5$ pair, degrades the combination to
$R = 1.3$--$1.5$ depending on reconstruction noise;
substituting the highest-redshift absolute BAO currently available
(Ly$\alpha$, $z = 2.33$) for the $z = 4.5$ pair, alongside the
full-$z$ lensing, gives $R = 1.4$--$1.5$.

\subsection{Step-like suppression of the matter power spectrum from massive neutrinos}

Beyond multi-probe cross-correlation analyses, the linear matter power
spectrum carries an intrinsic, DE-insensitive neutrino signature. Free
streaming sets a characteristic scale,
$k_{\rm nr} \approx 0.018\,\sqrt{\Omega_m\, m_\nu/{\rm eV}}\, h\,{\rm Mpc}^{-1}$
\citep{LESGOURGUES_2006, Hu_1998}, the minimum comoving free-streaming wavenumber,
attained near the non-relativistic transition
$z_{\rm nr} \approx 1890\, m_\nu/{\rm eV}$; this fixes the location
of the step deep in matter domination, where dark energy is negligible.
The depth then accumulates through the slowed CDM growth on
sub-free-streaming scales, $\delta_{\rm cdm} \propto a^{\,1-3f_\nu/5}$,
reaching the  asymptote $\Delta P/P \approx -8 f_\nu$ at $z=0$ \citep{Hu_1998}.

Most of this suppression is built up during matter domination: the depth accrues per $e$-fold of growth in proportion to the growth rate $f(a) \simeq \Omega_m(a)^{0.55}$~\citep{Wang_1998,PhysRevD.72.043529}, and the dark-energy era contributes only $\ln(1+z_{\rm DE}) \approx 0.5$ of the ${\sim}4$--$5$ $e$-folds accumulated since the non-relativistic transition, weighted by $f < 1$, hence $\lesssim 10\%$ of the depth. The remaining effect of late-time dark energy is a nearly scale-independent rescaling of the growth factor, $g(\Omega_m) \approx \Omega_m^{0.23}$ relative to matter domination~\citep{1992ARA&A..30..499C}, which cancels in the ratio $\Delta P/P$; the shape and present-day depth of the step therefore remain robust to late-time modifications of $H(z)$. A direct measurement of the step shape from upcoming spectroscopic surveys \citep{Schmittfull_2018} would thus provide a complementary, dark-energy-free probe of $\sum m_\nu$; we leave a quantitative forecast to
future work.

\section{Conclusions}
\label{sec:conclusions}

This work has examined whether current cosmological data can constrain
$\sum m_\nu$ independently of the assumed model of late-time dark
energy. Within $\Lambda$CDM, the combination of \textit{Planck} NPIPE
CMB, ACT~DR6 + SPT + \textit{Planck} lensing, and DESI~DR2 BAO yields
$\sum m_\nu < 0.056$~eV ($95\%$ CL), in $\sim 2$--$3\sigma$ tension with the
inverted-ordering floor.
This bound relaxes, however, as soon as the
dark-energy equation of state is freed.

We proceeded probe by probe, asking for each channel how much of its
$\sum m_\nu$ information survives once dark-energy assumptions are
relaxed. The unlensed primary CMB is robust; of the CMB lensing
information, only the four-point reconstruction $C_L^{\kappa\kappa}$
carries dark-energy-insensitive neutrino-mass information; and of the
BAO information, only the high-redshift distance shape survives. What
underlies all three is the total matter density: the data inform
$\sum m_\nu$ almost entirely through $\omega_m$
(Appendix~\ref{sec:anatomy}), whose anchor is the same low-redshift
distance scale which the dark-energy equation of state impacts.

We first pursue the conventional route, retaining all the data and
marginalizing over a parametrized expansion history rather than discarding
the dark-energy-sensitive information.
For $(w_0, w_a)$, this ``dark-energy-marginalized'' approach gives
$\sum m_\nu < 0.152$~eV (consistent with \cite{Elbers:2025vlz}).
Interestingly, the bound is
itself robust to further dark-energy freedom: binned and cubic $w(a)$
leave it within a few percent, with the additional expansion modes
resolved by the data (Section~\ref{sec:saturation}). 
This saturation
across the full sequence from constant $w$ to binned and cubic $w(a)$
has, to our knowledge, not previously been established (though see \cite{nair2025fourparamde} for a slightly different data combination); we trace it to
the additional dark-energy modes being nearly orthogonal to the matter
density $\omega_m$ through which late-time data inform $\sum m_\nu$
(Appendix~\ref{sec:anatomy}).
Hence $(w_0, w_a)$ marginalization thus suffices within smooth, matter-conserving dark energy with the standard sound speed $c_s^2 = 1$.

Assembling instead the channels robust to late-times into a single ``late-Universe-free''
combination, the CMB with $A_{\rm lens}$ marginalized (i.e., the ``unlensed'' CMB) and
$C_L^{\kappa\kappa}$ gives
$\sum m_\nu < 0.41$~eV with no parametrization of the late-time
expansion, a factor of $2.7$ weaker in $\sigma(\sum m_\nu)$ than the
marginalized bound. The bound is stable: it holds at $0.40$--$0.43$~eV on both
noiseless mock data and the real data, and across $\Lambda$CDM,
$(w_0, w_a)$, and binned and cubic $w(a)$, expressing no preference for
dark-energy freedom ($\Delta\chi^2 \approx 0$).

The two routes respond differently to the current data: the
dark-energy-marginalized bound tightens from $0.219$~eV on mock $\Lambda$CDM data
to $0.152$~eV on the real data, pulled by the same CMB--BAO distance
tension that drives the $(w_0, w_a)$ preference, whereas the
late-Universe-free bound is unchanged between the two, showing directly
that it does not draw on the disputed low-redshift information.

Given how tight the $\Lambda$CDM neutrino-mass bound has become, and its
sensitivity to the assumed expansion history, corroborating it with
measurements that relax different assumptions and draw on independent
probes becomes a priority: the full-shape galaxy clustering analyses of
current spectroscopic surveys~\cite{ivanov2026reanalyzingdesidr14,chudaykin2026reanalyzingdesidr12,Adame_2025}, and the
current and forthcoming $3\times2$pt and $6\times2$pt cross-correlations
of weak lensing and galaxy clustering.
These probes constrain the low-redshift expansion and growth through channels distinct from the geometric distances used here, and will be important to establishing a robust neutrino-mass measurement.

We note that our conservative cosmological bound $\sum m_\nu<0.41$ eV implies $m_{\nu_e}<0.134$ eV for individual neutrino state that is tested in laboratory settings. 
Thus, KATRIN, at its ultimate sensitivity of $m_{\nu_e}<0.3$ eV is not expected to see a positive signal of a nonzero neutrino mass.
A detection from KATRIN however would put into question our understanding of cosmology in  interesting ways.
On the other hand, Project 8 will have $m_{\nu_e}$ sensitivity comparable to the conservative cosmological bounds and is, therefore, highly complementary. 

Looking ahead, the marginalized bound sharpens to
$\sigma(\sum m_\nu) \approx 43$~meV with five-year Simons
Observatory lensing and Spec-S5 BAO, with $(w_0, w_a)$ marginalization
remaining saturated, while the late-Universe-free combination retains its
robustness at forecast precision, tightening from $0.41$ to $0.31$~eV
and to $0.28$~eV at the cosmic-variance limit, with $R \approx 1$
throughout (Section~\ref{sec:forecast}). Galaxy surveys offer no
equally clean late-Universe-free channel, but under $(w_0, w_a)$
marginalization the LSST + Simons Observatory $6\times2$-point
combination reaches $\sigma(\sum m_\nu) = 60$~meV with the
mildest degradation of the combinations we forecast ($R = 1.4$),
without using BAO. Finally, a direct measurement of the step-like
suppression that massive neutrinos imprint on the matter power
spectrum, whose shape and depth are fixed deep in matter domination,
would provide a dark-energy-free signature of $\sum m_\nu$ itself
(Section~\ref{sec:forecast}).

\section*{Acknowledgements}
Work done at SLAC National Accelerator Laboratory was supported by the Department of Energy under contract DE-AC02-76SF00515. WLKW acknowledges support from an Early Career Research Award DE-SC0026376 of the Department of Energy. We thank Noah Sailer, Matthew Johnson, Santiago Agui, Irene Abril-Cabezas, Abhishek Maniyar, Zachary Weiner, Kendrick Smith, Wayne Hu, Jessica Zebrowski, Tanvi Karwal, Scott Dodelson, Marilena Loverde, Misha Ivanov, Neal Dalal and Mathew Madhavacheril for useful discussions. FJQ is grateful to xc340 for all the memories shared. Computations were performed on the Niagara supercomputer at the SciNet HPC Consortium. SciNet is funded by Innovation, Science and Economic Development Canada; the Digital Research Alliance of Canada; the Ontario Research Fund: Research Excellence; and the University of Toronto. This work was completed at the Aspen Center for Physics, which is supported by National Science Foundation grant PHY-2210452.

\bibliography{apssamp}

\appendix

\section{Implementation and validation of the $\alpha_{\rm DE}$ parameter}
\label{app:alpha_DE}

\subsection{CAMB modification}

We implement $\alpha_{\rm DE}$ by modifying the CAMB Boltzmann code~\citep{Lewis:1999bs} to rescale the expansion rate at low redshift during the integration of the photon-baryon perturbations. Specifically, we add a new Fortran module (\texttt{Hz\_template.f90}) that loads a precomputed $H(z)$ template evaluated at the Planck $\Lambda$CDM best-fit cosmology on a grid of 1001 points from $z = 0$ to $z = 10$. During the Boltzmann integration, the conformal time integrand $dt/da \propto 1/(a^2 H)$ is replaced by $1/(a^2 \, \alpha_{\rm DE} \, H_{\rm template})$ for $z < z_{\rm DE} = 0.706$, with the true cosmological $H(z)$ used at higher redshifts.

A critical design requirement is that $\alpha_{\rm DE}$ must affect only the primary CMB anisotropies computed through the Boltzmann hierarchy, and not the CMB lensing potential $C_L^{\phi\phi}$ or any distance observables used by external likelihoods (e.g., BAO). We achieve this through a dual-CAMB architecture: one instance of CAMB with $\alpha_{\rm DE}$ enabled computes the unlensed $C_\ell^{TT, TE, EE}$, while a second, unmodified instance computes the lensing potential, all distances, and the sound horizon $r_d$. The lensed CMB spectra are then constructed by applying the standard lensing operation to the $\alpha_{\rm DE}$-modified unlensed spectra using the lensing potential from the unmodified instance. This ensures that the lensing power spectrum and BAO observables respond to the true cosmological parameters (including $w_0$, $w_a$) independently of $\alpha_{\rm DE}$.

\subsection{Validation}

\begin{figure}[hbt!]
\centering
\includegraphics[width=0.8\columnwidth]{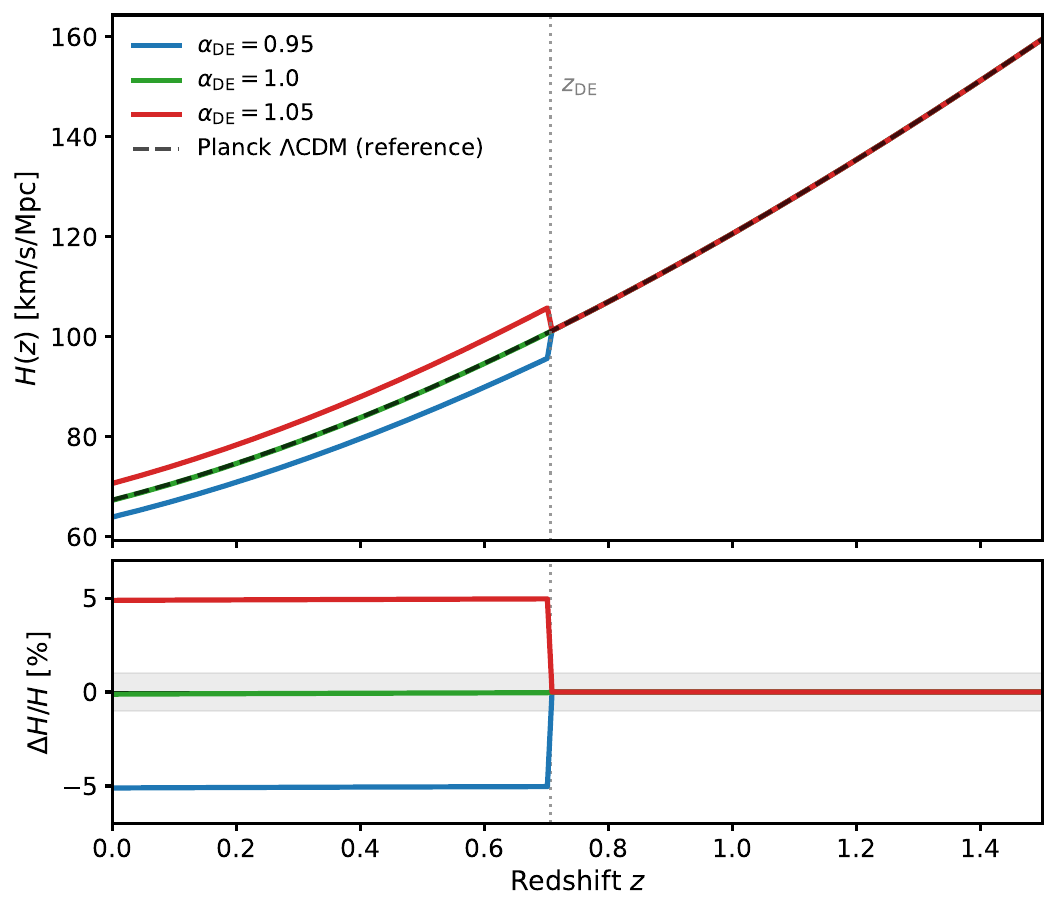}
\includegraphics[width=0.8\columnwidth]{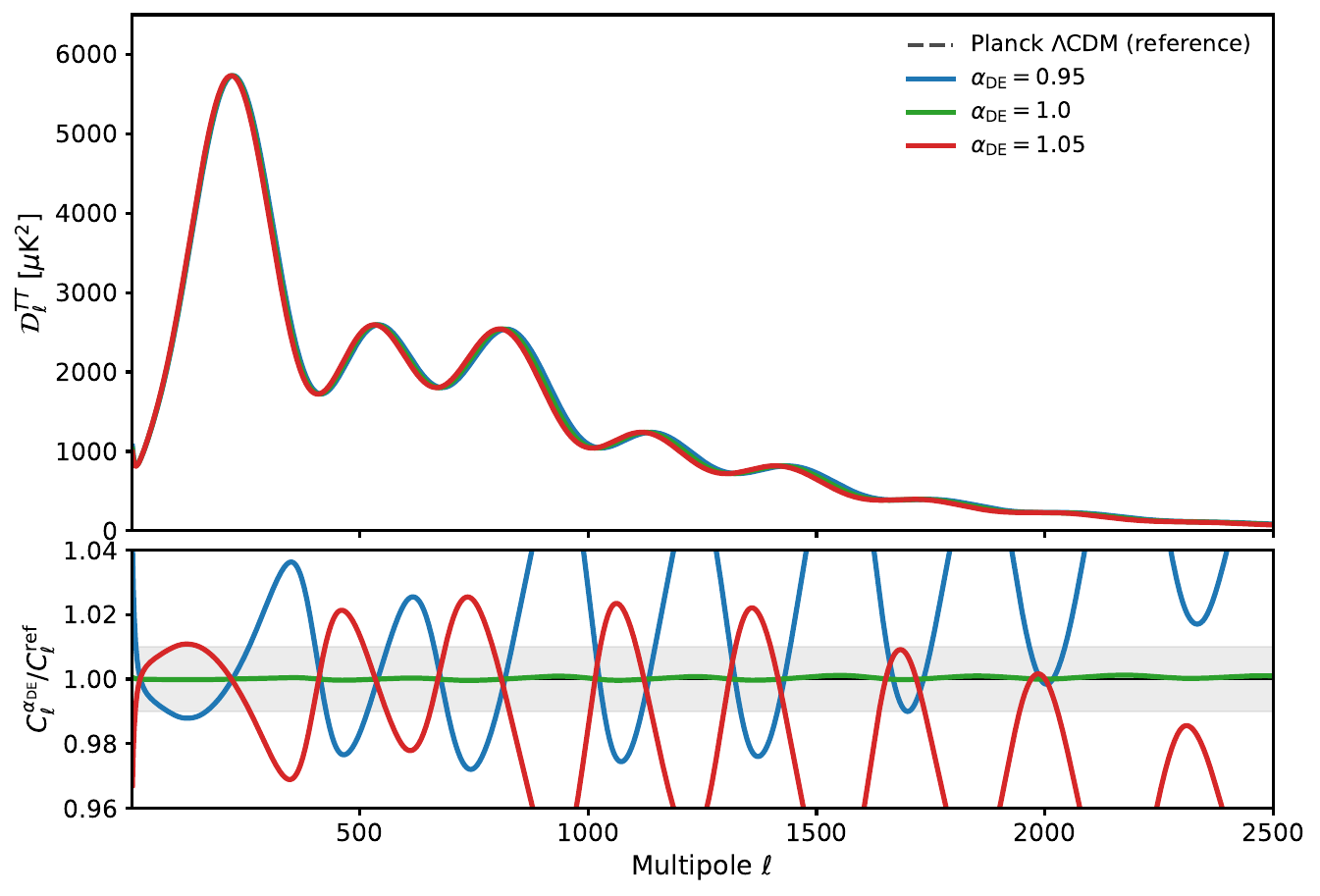}
    \caption{Effect of the $\alpha_\mathrm{DE}$ parameter on the expansion history and CMB. \textit{Top figure}: The expansion rate $H(z)$ (upper panel) and fractional difference $\Delta H/H$ (lower panel) for $\alpha_\mathrm{DE} = 0.95$, $1.0$, and $1.05$. For $z < z_\mathrm{DE} = 0.706$, $H(z)$ is rescaled by $\alpha_\mathrm{DE}$ relative to the Planck $\Lambda$CDM template; for $z > z_\mathrm{DE}$, all curves converge to the true cosmology. \textit{Bottom figure}: The CMB temperature power spectrum $\mathcal{D}_\ell^{TT}$ (upper panel) and ratio to the reference spectrum (lower panel). A $\pm 5\%$ change in $\alpha_\mathrm{DE}$ produces $2$--$4\%$ oscillatory deviations at the acoustic peaks, with a characteristic shift in peak positions toward higher (lower) $\ell$ for $\alpha_\mathrm{DE} < 1$ ($\alpha_\mathrm{DE} > 1$). This behavior is consistent with the modified late-time expansion history altering the angular diameter distance to recombination $D_A(z_*)$: smaller $\alpha_\mathrm{DE}$ increases the integrated distance, shifting the peaks to smaller angular scales (higher $\ell$), while larger $\alpha_\mathrm{DE}$ has the opposite effect. The peak shifts shown here are computed at fixed $H_0$ (and hence fixed $r_\star$),
which isolates the bare $\alpha_{\rm DE}$ imprint on $D_A(z_\star)$; in the MCMC the
acoustic scale $\theta_\star$ is a sampled, data-constrained quantity, so this shift is
reabsorbed into the inferred $H_0$ and the fitted peak positions remain fixed.
}

\label{fig:alpha_DE_diagnostic}
\end{figure}

We perform the following checks to validate the implementation:

\begin{enumerate}
    \item \textit{Internal consistency}: Setting $\alpha_{\rm DE} = 1.0$ recovers the standard CAMB $C_\ell$ to within $0.1$--$0.2\%$, consistent with the residual interpolation error from the discrete $H(z)$ template grid. The corresponding shift in high-$\ell$ CMB $\chi^2$ is $\lesssim 3.4$ units, which is negligible compared to the $> 1000$ unit shift induced by a $5\%$ change in $\alpha_{\rm DE}$.

    \item \textit{Lensing isolation}: Varying $\alpha_{\rm DE}$ by $\pm 5\%$ in the dual-CAMB setup produces identical $C_L^{\phi\phi}$ to machine precision, confirming that the lensing potential is computed entirely from the unmodified CAMB instance.

    \item \textit{Distance isolation}: The sound horizon $r_d$, angular diameter distances, and $H(z)$ values returned to BAO likelihoods are unaffected by $\alpha_{\rm DE}$, as verified by comparing outputs with $\alpha_{\rm DE}$ enabled and disabled.

    \item \textit{CMB effect}: The impact of $\alpha_{\rm DE}$ on the temperature power spectrum is shown in Fig.~\ref{fig:alpha_DE_diagnostic}. A $\pm 5\%$ change produces $2$--$4\%$ oscillatory deviations at the acoustic peaks, with peak positions shifting toward higher (lower) $\ell$ for $\alpha_{\rm DE} < 1$ ($> 1$), consistent with the expected change in $D_A(z_*)$.
\end{enumerate}

\subsection{Parameter choices}

We adopt a uniform prior $\alpha_{\rm DE} \in [0.8, 1.2]$ with a Gaussian proposal centered at $1.0$ with width $0.01$. The reference redshift $z_{\rm DE} = 0.706$ matches the reference redshift used for the BAO distance differences (Section~\ref{sec:bao}). We have verified that the posterior on $\alpha_{\rm DE}$ is well contained within the prior: in $\Lambda$CDM, $\alpha_{\rm DE} = 0.997 \pm 0.015$, and in $w_0 w_a$CDM, $\alpha_{\rm DE} = 1.056 \pm 0.036$.

\section{Computation of the two-point and four-point lensing Fisher information}
\label{app:response_kernel}

This appendix describes how we compute the Fisher information on $\sum m_\nu$ carried by the
two-point (peak smoothing) and four-point (reconstructed $C_L^{\kappa\kappa}$) lensing
channels, decomposed by lensing multipole $L$ and source redshift $z$
(Figure~\ref{fig:2pt_4pt_decomposition}).

\subsubsection*{Four-point channel}

The four-point channel measures $C_L^{\kappa\kappa}$ directly at each multipole $L$. Its
Fisher information on a parameter $\theta$ is
\begin{equation}
    F_{4\mathrm{pt}}(\theta) = \sum_L
    \frac{[\partial C_L^{\kappa\kappa} / \partial\theta]^2}{\sigma_L^2}\,,
\end{equation}
where $\sigma_L$ are the combined ACT DR6 + SPT + \textit{Planck} reconstruction
uncertainties (19 bandpower bins, interpolated with a cubic spline). Since each multipole
$L$ is measured independently, the Fisher density $dF/dL$ is exactly additive in $L$. To
decompose by source redshift, we replace $C_L^{\kappa\kappa}$ with its cumulative
contribution from $z < z_{\rm cut}$,
\begin{equation}
    C_L^{\kappa\kappa}(z < z_{\rm cut})
    = \int_0^{z_{\rm cut}} dz\;\frac{dC_L^{\kappa\kappa}}{dz}\,,
\end{equation}
computed from the lensing kernel output of \texttt{class\_sz}~\citep{Bolliet:2023eob}. Derivatives
with respect to $\sum m_\nu$ are evaluated via centred finite differences at five
cosmologies ($\sum m_\nu^{\rm fid} \pm \delta$, $w^{\rm fid} \pm \delta$, and the
fiducial), holding the unlensed CMB fixed. The cumulative kernel is evaluated on a grid of
$300$ lensing multipoles ($20 \le L \le 3000$) and $600$ redshift values
($0.01 \le z \le 50$).

\subsubsection*{Two-point channel: response kernel}

The two-point lensing effect enters through the non-perturbative convolution that produces
the lensed $\tilde{C}_\ell^{TT}$ from the unlensed $C_\ell^{TT}$ and the lensing potential
$C_L^{\phi\phi}$. Unlike the four-point case, a change in $C_L^{\phi\phi}$ at a single
multipole $L$ affects $\tilde{C}_\ell^{TT}$ at all CMB multipoles $\ell$ simultaneously, so
the Fisher information cannot be trivially decomposed by $L$.

To map which lensing scales contribute, we compute the response kernel
\begin{equation}
    K_{\ell L} \equiv \frac{\partial \tilde{C}_\ell^{TT}}{\partial C_L^{\phi\phi}}\,,
\end{equation}
which encodes how a perturbation to the lensing potential at multipole $L$ propagates to the
lensed CMB at multipole $\ell$. We evaluate $K_{\ell L}$ numerically at $50$ values of $L$
spanning $20 \le L \le 2000$. For each $L_i$, we add a narrow Gaussian bump to the
fiducial $C_L^{\phi\phi}$,
\begin{equation}
    C_L^{\phi\phi,\mathrm{bumped}} = C_L^{\phi\phi,\mathrm{fid}}
    + A\,\exp\!\left[-\frac{(L - L_i)^2}{2\sigma_i^2}\right],
\end{equation}
where $A = 0.01 \times C_{L_i}^{\phi\phi,\mathrm{fid}}$ (a $1\%$ perturbation, well within
the linear regime) and $\sigma_i = \max(\Delta L_i / 2,\; 3)$ with $\Delta L_i$ the local
grid spacing. The bump is constructed on a fine integer-$L$ grid to ensure it is
well-sampled even at low $L$. We then run the full non-perturbative correlation function
lensing pipeline following the method of \citep{Lewis:1999bs}:
\begin{enumerate}
    \item Compute the $\theta$-dependent lensing quantities $\sigma^2(\theta)$ and
          $C_{g\ell,2}(\theta)$ from $C_L^{\phi\phi}$ via Wigner $d$-function sums.
    \item Evaluate the lensing correction
          $\Delta\xi(\theta) = \tilde{\xi}(\theta) - \xi(\theta)$ using the exponential
          damping factor $e^{-\ell(\ell+1)\sigma^2(\theta)/2}$ and the gradient--lensing
          coupling $C_{g\ell,2}(\theta)$.
    \item Invert to harmonic space via Legendre transform on a uniform $\theta$-grid
          with Gaussian apodisation near $\theta = \pi$.
\end{enumerate}
The response kernel is then
\begin{equation}
    K_{\ell L_i} = \frac{\tilde{C}_\ell[C_L^{\phi\phi} + \mathrm{bump}]
    - \tilde{C}_\ell[C_L^{\phi\phi,\mathrm{fid}}]}
    {\int \mathrm{bump}(L)\,dL}\,.
\end{equation}
We have verified this pipeline against the independent lensing implementation in
\texttt{class\_sz}, finding agreement at the $\sim 3\%$ level for $\ell \le 2500$.

\subsubsection*{ Fisher density}

Given the response kernel, the derivative of the lensed spectrum with respect to $\theta$ is
a coherent sum over $L$:
\begin{equation}
    \frac{\partial \tilde{C}_\ell^{TT}}{\partial\theta}
    = \sum_L K_{\ell L}\,\frac{\partial C_L^{\phi\phi}}{\partial\theta}\,\Delta L\,.
\end{equation}
Because the two-point Fisher involves squaring this sum, one cannot cleanly assign Fisher
information to individual $L$. We therefore define an approximate Fisher density per $L$ by
treating each $L$ as if it contributed independently:
\begin{equation}
\label{eq:AL}
    A(L_i;\, z < z_{\rm cut}) = \sum_\ell
    \frac{\bigl[K_{\ell L_i}\;\partial C_{L_i}^{\phi\phi}(z < z_{\rm cut})
    / \partial\theta\bigr]^2}
    {\mathrm{Var}(\tilde{C}_\ell^{TT})}\;\Delta L_i\,,
\end{equation}
where $\mathrm{Var}(\tilde{C}_\ell^{TT}) = 2(\tilde{C}_\ell + N_\ell)^2
/ [(2\ell+1)f_{\rm sky}]$ and $N_\ell$ is the instrumental noise power spectrum. For the
baseline results we adopt \textit{Planck}-like noise ($40\,\mu$K$\cdot$arcmin, $5'$ beam,
$f_{\rm sky} = 0.7$); Appendix~\ref{app:2pt_noise_robustness} shows that the shape of
$A(L)$ is insensitive to the noise model.

The total $\sum_L A(L)$ does not equal the true two-point Fisher (which involves cross-$L$
interference from the coherent sum), but the shape of $A(L)$ correctly identifies the
dominant $L$-range. This is confirmed by an independent $L_{\max}$ truncation scan, in
which we run the full lensing pipeline with $C_L^{\phi\phi}$ set to zero for $L > L_{\max}$
and compute the resulting Fisher at each truncation; the cumulative fractions are consistent
with those obtained from $A(L)$.

To obtain the redshift decomposition (Figure~\ref{fig:2pt_4pt_decomposition}, bottom panel),
we sum $A(L)$ over all $L$ at each cumulative redshift cut $z_{\rm cut}$, differentiate the
resulting cumulative Fisher with respect to $z$, and normalise to unit area.

\section{Redshift weighting of two-point lensing sensitivity to neutrino mass: robustness to experimental noise}
\label{app:2pt_noise_robustness}

\begin{figure}[h]
\centering
\includegraphics[width=\columnwidth]{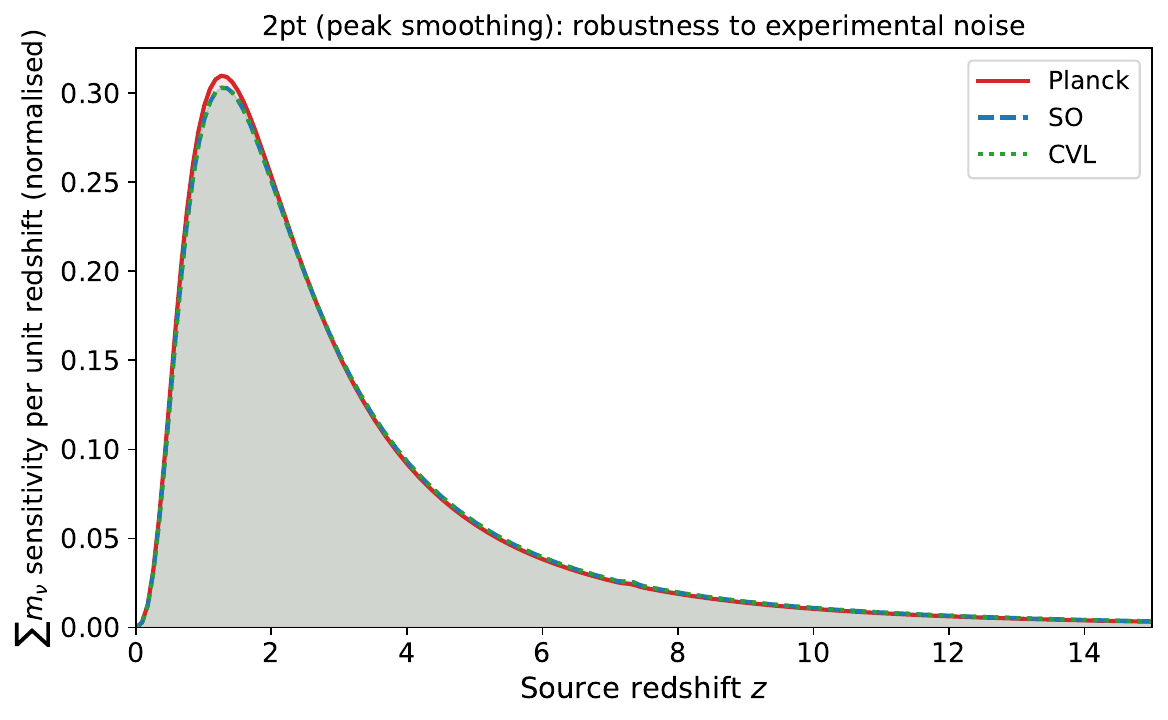}
\caption{Two-point $\sum m_\nu$ sensitivity per unit source redshift, normalised to unit area, for three experimental noise models. The curves are nearly indistinguishable, confirming that the redshift weighting of the two-point channel is a geometric property of the lensing convolution rather than a feature of the instrumental noise.}
\label{fig:2pt_experiments}
\end{figure}

The redshift distribution of the two-point lensing sensitivity is set primarily by the geometry of the lensing convolution, not by the instrumental noise. To demonstrate this, we recompute the two-point $\sum m_\nu$ sensitivity per unit redshift (as in Figure~\ref{fig:2pt_experiments}) for three noise models: \textit{Planck} ($40\,\mu$K$\cdot$arcmin, $5'$ beam, $f_{\rm sky} = 0.7$), the Simons Observatory ($6\,\mu$K$\cdot$arcmin, $1.4'$ beam, $f_{\rm sky} = 0.4$), and the cosmic variance limit ($f_{\rm sky} = 1$). In all three cases we hold the fiducial lensing power spectrum and response kernel $K_{\ell L}$ fixed; only the noise contribution to ${\rm Var}(\tilde{C}_\ell^{TT})$ changes.

Figure~\ref{fig:2pt_experiments} shows the result. The three curves are nearly indistinguishable. Table~\ref{tab:2pt_experiments} confirms this quantitatively: the peak sensitivity remains at $z \approx 1.3$ for all three experiments, the median redshift shifts by less than $0.1$ (from $2.35$ for \textit{Planck} to $2.43$ for the cosmic variance limit), and the fraction of sensitivity originating from $z < 1$ changes by less than half a percentage point. The Simons Observatory is already effectively at the cosmic variance limit for this diagnostic.

\begin{table}[h]
\centering
\caption{Two-point lensing sensitivity to $\sum m_\nu$ as a function of source redshift, for three experimental configurations. The peak and median redshifts, as well as the cumulative fractions from low redshift, are nearly identical across all three cases.}
\label{tab:2pt_experiments}
\begin{tabular}{lcccc}
\toprule
Experiment & Peak $z$ & Median $z$ & $z < 1$ & $z < 2$ \\
\midrule
\textit{Planck} & 1.26 & 2.35 & 13.6\% & 42.9\% \\
SO & 1.26 & 2.43 & 13.2\% & 41.9\% \\
CVL & 1.26 & 2.43 & 13.2\% & 41.9\% \\
\bottomrule
\end{tabular}
\end{table}

This insensitivity arises because the two-point $L$-weighting is dominated by the response kernel $K_{\ell L}$, which encodes how lensing modes at multipole $L$ couple to CMB multipoles $\ell$ through the non-perturbative correlation function (see Appendix~\ref{app:response_kernel}). This coupling is a geometric property of the lensing remapping and does not depend on the noise. Changing the noise model reweights which CMB multipoles $\ell$ contribute most to the Fisher sum, but the $L$-weighting (and hence the redshift weighting) is already saturated by $\ell \lesssim 2500$, where even \textit{Planck} is signal-dominated.

\section{Mock primary CMB spectra}
\label{app:fiducial_cmb}

Throughout the mock-data sections of this work we use mock data vectors for the CMB lensing, BAO, and primary CMB high-$\ell$ (CamSpec TT+TE+EE) likelihoods, generated at the Planck NPIPE $\Lambda$CDM best-fit cosmology ($\omega_b = 0.02236$, $\omega_c = 0.12008$, $\theta_{\rm MC} = 0.01041$, $n_s = 0.9650$, $A_s = 2.113\times10^{-9}$, $\tau = 0.0571$, $H_0 = 67.28$~km/s/Mpc, $\sum m_\nu = 0.06$~eV), while retaining the real low-$\ell$ likelihoods (Section~\ref{sec:metric}). This choice ensures that our results are not affected by statistical fluctuations in the data, isolating the information content of each channel in a controlled setting. The mock replacement of the primary CMB requires particular justification: the real Planck primary CMB data introduces a bias in the two-point lensing channel, which this appendix describes and quantifies.

It is well known that the Planck temperature and polarization power spectra prefer a lensing amplitude slightly above unity; the NPIPE CamSpec likelihood finds $A_{\rm lens} = 1.095 \pm 0.056$~\citep{Rosenberg_2022}. This excess smoothing of the acoustic peaks enhances the sensitivity of the primary CMB to parameters that affect lensing, including $\sum m_\nu$. When we use the real Planck spectra (which contain this fluctuation), the two-point lensing constraint on $\sum m_\nu$ is $0.36$~eV ($\Lambda$CDM, 95\% CL); this is tighter than the four-point constraint of $0.41$~eV, giving the impression that two-point lensing is the more powerful channel. To test whether this hierarchy is physical or an artifact of the $A_{\rm lens}$ fluctuation, we generate a mock primary CMB data vector at $A_{\rm lens} = 1$ and repeat the analysis.

We construct a mock replacement for the CamSpec NPIPE high-$\ell$ TT+TE+EE data vector as follows. We compute the theory $C_\ell^{TT}$, $C_\ell^{TE}$, and $C_\ell^{EE}$ at the Planck NPIPE $\Lambda$CDM best-fit cosmology using CAMB, with $A_{\rm lens} = 1$ (the physical value). For each of the four TT frequency cross-spectra ($100\times100$, $143\times143$, $217\times217$, $143\times217$), we add the best-fit foreground model (power-law residuals evaluated at the best-fit nuisance parameters from our MCMC chains) and divide by the best-fit calibration factors:
\begin{equation}
\begin{gathered}
\hat{C}_\ell^{(i)} = \frac{C_\ell^{TT} + C_\ell^{{\rm fg},(i)}}{c^{(i)}}, \\[2pt]
i \in \{100\!\times\!100,\ 143\!\times\!143,\ 217\!\times\!217,\ 143\!\times\!217\}.
\end{gathered}
\end{equation}
and similarly $\hat{C}_\ell^{TE} = C_\ell^{TE}/c^{TE}$ and $\hat{C}_\ell^{EE} = C_\ell^{EE}/c^{EE}$, where $c^{(i)}$ are the CamSpec calibration factors. The resulting six-column array replaces the \texttt{cl\_hat\_file} in the CamSpec dataset configuration, while the covariance matrix, foreground templates, and all other likelihood infrastructure remain unchanged. We verify that evaluating the likelihood at the best-fit parameters yields $\chi^2 \approx 0$ with the mock data vector.

\begin{figure}
\centering
\includegraphics[width=\columnwidth]{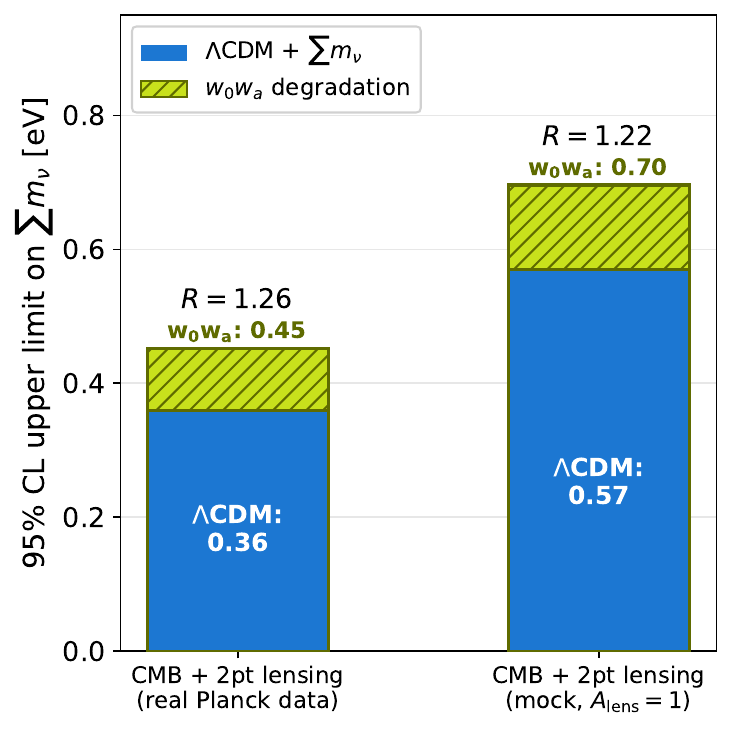}
\caption{Effect of the Planck $A_{\rm lens}$ fluctuation on the two-point lensing constraint. Left: real Planck NPIPE primary CMB data; right: mock primary CMB at $A_{\rm lens} = 1$. Both use mock CMB lensing data. The two-point constraint weakens by $\sim$60\% when the $A_{\rm lens} > 1$ fluctuation is removed, while the ratio $R$ remains similar. The four-point constraint (not shown) is unaffected.}
\label{fig:camspec_fiducial_bar}
\end{figure}

Figure~\ref{fig:camspec_fiducial_bar} compares the two-point lensing constraint on $\sum m_\nu$ using the real Planck spectra and the mock spectra. With mock data ($A_{\rm lens} = 1$), the $\Lambda$CDM constraint weakens from $0.36$~eV to $0.57$~eV, a degradation of approximately 60\%. The $w_0 w_a$CDM constraint weakens comparably, from $0.45$~eV to $0.70$~eV, while the ratio $R$ remains similar ($1.26 \to 1.22$). The four-point constraint, by contrast, is unchanged ($0.41$~eV with either dataset), as expected since $A_{\rm lens}$ marginalizes over the two-point channel and the external $C_L^{\kappa\kappa}$ data is fiducial in both cases.

This confirms that the apparent advantage of two-point over four-point lensing in the real data is an artifact of the Planck $A_{\rm lens}$ fluctuation. On unfluctuated data, the four-point channel ($0.41$~eV) provides a tighter constraint than the two-point channel ($0.57$~eV); the mock-data sections of the main text therefore adopt the mock primary CMB spectra alongside the mock lensing and BAO vectors. The real-data results of Section~\ref{sec:combined}, which necessarily retain the fluctuation, are correspondingly conservative: the two-point lensing contribution to the standard combination is enhanced by it, making the gap between the standard and late-Universe-free bounds appear smaller than it would be on average.

\begin{figure}[t]
    \centering
    \includegraphics[width=\columnwidth]{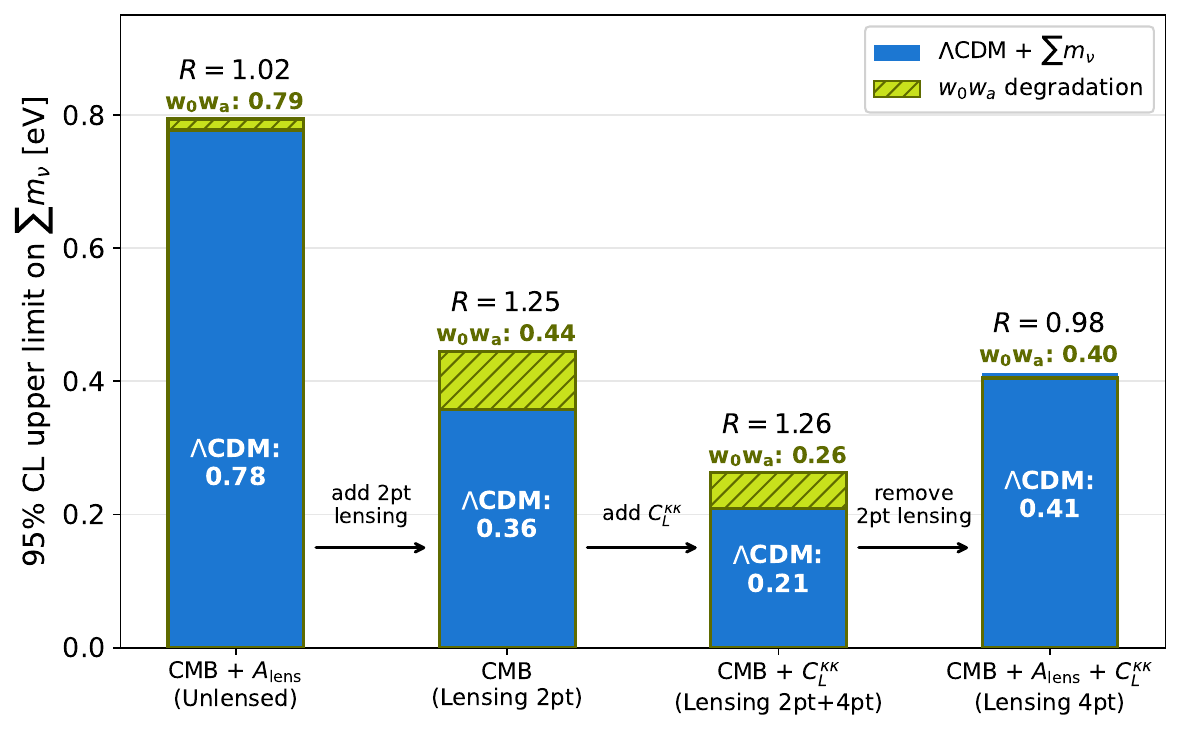}
    \caption{Same as Figure~\ref{fig:lensed_cmb} but using real Planck NPIPE primary CMB data instead of mock spectra. The two-point lensing constraint ($0.36$~eV) is substantially tighter than in the mock case ($0.57$~eV), reflecting the well-known Planck preference for $A_{\rm lens} \approx 1.10$. This excess peak smoothing enhances the sensitivity of the two-point channel to $\sum m_\nu$, making it appear more constraining than the four-point channel ($0.41$~eV). The four-point constraint and its dark energy insensitivity ($R = 0.98$) are unchanged, as expected since $A_{\rm lens}$ marginalizes over the two-point contribution.}
    \label{fig:lensed_cmb_real}
\end{figure}

\section{Anatomy of the constraint: the $\omega_m$ channel}
\label{sec:anatomy}

Late-time data inform $\sum m_\nu$ through a single quantity, the physical matter density $\omega_m$. With $\omega_\nu = \Omega_\nu h^2 = \sum m_\nu/(93.14~\mathrm{eV})$ and, in a flat universe, $\omega_m = \omega_b + \omega_c + \omega_\nu$, the mass sum is the residual of the matter budget,
\begin{equation}
\frac{\sum m_\nu}{93.14~\mathrm{eV}} = \omega_m - \omega_{cb},
\qquad \omega_{cb} \equiv \omega_b + \omega_c.
\label{eq:nu_budget}
\end{equation}
The CMB fixes the cold density $\omega_{cb}$ to better than a percent from the acoustic-peak morphology, early-universe physics independent of the late-time expansion. Any late-time probe therefore constrains $\sum m_\nu$ only through $\omega_m$: it must sharpen the total matter density against the CMB-fixed cold component~\cite{Loverde:2024nfi,Lynch:2025ine}.

The dominant late-time constraint on $\omega_m$ comes from the BAO ruler. A survey at redshift $z$ measures the ratios $D_M(z)/r_d$ and $D_H(z)/r_d$, where $D_H(z) = c/H(z)$ and the drag-epoch sound horizon $r_d = f(\omega_b, \omega_{cb})$ is set by pre-recombination physics, independent of $w(a)$. Here $r_d$ depends on the cold physical density $\omega_{cb} = \omega_b + \omega_c$, not on $\omega_m$, because neutrinos are still relativistic at the drag epoch $z_{\rm drag} \approx 1060$; the residual dependence on $\omega_\nu$ is below one percent~\citep{Aubourg_2015}. Although each measurement is a distance ratio, it still constrains the densities through $r_d$, because $r_d$ is not a free normalization but a fixed function of $\omega_b$ and $\omega_{cb}$. At redshifts above $z_{\rm DE}$, where dark energy is negligible, the expansion rate is $H(z) \simeq 100\sqrt{\omega_m}\,(1+z)^{3/2}~\mathrm{km\,s^{-1}\,Mpc^{-1}}$, independent of $h$ and of $w(a)$, so the radial ratio satisfies $D_H(z)/r_d \propto 1/[\sqrt{\omega_m}\,r_d(\omega_b, \omega_{cb})]$. A BAO measurement at $z > z_{\rm DE}$ therefore determines the combination $\sqrt{\omega_m}\,r_d$ with no assumption about dark energy, and this is the dark-energy-robust part of the constraint. The CMB adds one more point on the same ruler through the acoustic scale $\theta_* = r_s(z_*)/D_M(z_*)$ at $z_* \approx 1100$. The quantity $D_M(z_*)$ integrates the expansion history through the dark-energy era, so $\theta_*$ alone does not fix $\omega_m$ once $w(a)$ is free. What is dark-energy-robust is the distance interval between the BAO redshifts and $z_*$: this interval is accumulated almost entirely at $z > z_{\rm DE}$, so combining the BAO ratios with $\theta_*$ extends the redshift range over which the dark-energy-robust combination $\sqrt{\omega_m}\,r_d$ is measured, and thereby tightens $\omega_m$~\cite{Weiner:2026sfm, Loverde:2024nfi, Lynch:2025ine}. Equivalently, the dark-energy-robust information is the value of $\sqrt{\omega_m}\,r_d$; it can be read from any single BAO measurement at $z > z_{\rm DE}$, and the CMB point at $z_*$ lengthens the matter-era lever arm over which it is determined.
 The remaining, dark-energy-sensitive part of the ruler is the relative growth of $D_M$ across the low-redshift survey range, which is governed by the ratio of dark-energy to matter density. Computed with \textsc{camb} distances at fixed $\theta_*$, a $+3\%$ change in $\omega_m$ shifts $D_M(z)/r_d$ by the same coherent amount across all BAO redshifts, whereas a change in $w_0$ tilts only the low-redshift part.

Two operations remove dark-energy sensitivity, with opposite consequences for $\omega_m$. Marginalizing $w(a)$ projects out the shape directions; these are only partially aligned with the anchor, and the CMB limits the allowed $\omega_m$ range to $\sim 0.8\%$ regardless of $w(a)$, so the matter-density information is degraded but not lost. The marginal $\sigma(\omega_m)$ moves from $0.0006$ in $\Lambda$CDM to $0.0011$ under $(w_0, w_a)$, short of the no-BAO value $0.0019$; adding binned or cubic $w(a)$ removes only near-orthogonal modes and leaves $\sigma(\omega_m)$ near $0.0011$. This is the origin, at the level of the matter density, of the bound saturation of Section~\ref{sec:saturation}.

Differencing the distances, $\Delta D_M = D_M(z) - D_M(z_{\rm ref})$, removes most of the constraint on $\omega_m$ but makes the resulting bound insensitive to the dark-energy model, and the two facts have the same origin. At fixed $\theta_*$, raising $\omega_m$ shifts $D_M(z)/r_d$ upward by a nearly common amount across the BAO range: over the DESI redshifts the response varies by only about $25\%$ about its mean, so to good approximation it is a redshift-independent offset rather than a change of shape. A change in $w(a)$, by contrast, tilts $D_M(z)/r_d$: its effect grows with redshift across the survey range, varying by of order a factor of two. In the absolute distances these two responses are nearly parallel as functions of redshift (their shapes overlap at the $95\%$ level over the DESI tracers), which is the geometric statement of the $\omega_m$--$w(a)$ degeneracy: absolute BAO cannot separate a shift in the matter density from a change in the expansion history, so freeing $w(a)$ corrupts its determination of $\omega_m$ and hence of $\sum m_\nu$. Differencing subtracts the common offset, cancelling most of the $\omega_m$ response while leaving the $w(a)$ tilt largely intact: over the BAO range the $\omega_m$ leverage of $D_M/r_d$ drops by an order of magnitude, whereas the $w(a)$ leverage drops by only a factor of a few. The combination $\sqrt{\omega_m}\,r_d$ still enters $\Delta D_M/r_d$, but with so little leverage that the differenced observable is dominated by the part of the response that tracks $w(a)$ and no longer carries appreciable matter-density information. This is why the differenced bound is dark-energy-robust ($R \approx 1$; Section~\ref{sec:bao}): with the $\omega_m$ leverage gutted, marginalizing over the dark-energy model removes almost nothing, so the bound does not move with the assumed $w(a)$. It is robust precisely because it is weak: differencing keeps the dark-energy-sensitive shape but discards the coherent offset through which BAO would otherwise pin $\omega_m$, so $\Delta D_M$ adds no $\omega_m$ information beyond the CMB and lensing determination. Note that $r_d$ itself is unchanged by differencing; but $r_d$ is a derived function of $\omega_b$ and $\omega_{cb}$ rather than a measured quantity, so leaving it in place carries no information on its own. Differencing is thus the conservative one of the two routes to a late-Universe-free bound in Section~\ref{sec:combined}: it trades constraining power for insensitivity to the expansion history.

 Differencing helps only against non-background dark energy, dark-energy clustering, modified gravity, or an $r_d$ miscalibration, which shift the overall amplitude that distance ratios cancel; against smooth $w(a)$ it only weakens the bound.

These statements are quantitative. Across ten mock-data chains spanning a threefold range in precision and covering $\Lambda$CDM, $w$CDM, $(w_0, w_a)$, and binned and cubic $w(a)$ with all-BAO ($D_M+D_H$), differenced ($\Delta D_M$ and $\Delta D_M+D_H$), and no-BAO distance treatments, the marginal widths obey
\begin{equation}
\sigma\!\left(\sum m_\nu\right) \approx k\,\sigma(\omega_m), \qquad k \approx 63.1~\mathrm{eV},
\label{eq:lockstep}
\end{equation}
to within $6\%$ (Figure~\ref{fig:omegam_lockstep}). The coefficient is $k \approx 63.1$~eV rather than the $93.14$~eV of Eq.~\eqref{eq:nu_budget} because $\sum m_\nu$ is the difference of two comparable-uncertainty densities, $\omega_m$ from the BAO anchor and $\omega_{cb}$ from the CMB, positively correlated at $r \approx 0.7$--$0.8$, which narrows the difference by $\sim 1.5\times$. The relation identifies $\omega_m$ as the channel by which late-time data inform $\sum m_\nu$, and the dark-energy sensitivity of a probe as the degree to which its determination of $\omega_m$ collapses once the expansion history is freed. It applies to the distance contribution in the late-Universe-free or dark-energy-marginalized regime; a residual free-streaming bound $\sum m_\nu \lesssim 0.4$~eV remains from the CMB and four-point lensing with no distance information at all (Section~\ref{sec:combined}).

\section{Loss of matter-density sensitivity in BAO distance differences}
\label{sec:bao_diff_omegam}

We provide intuition for why the differencing approach
of Section~\ref{sec:bao} reduces the constraining power of the transverse
BAO distances, and we clarify whether the matter density or the physical matter density is degraded.
This distinction matters: the neutrino mass enters cosmology through the
physical density $\omega_m \equiv \Omega_m h^2$ via
$\omega_\nu = \omega_m - \omega_b - \omega_c$, so $\sigma(\Sigma m_\nu)$
is controlled by $\sigma(\omega_m)$, not by $\sigma(\Omega_m)$. We will
see that differencing produces two logically distinct degradations: a
$\sim 2\times$ loss of the $\omega_m$ amplitude precision (which is the
quantity relevant to $\Sigma m_\nu$), and a separate, larger broadening
of $\sigma(\Omega_m)$ and $\sigma(H_0)$ along the $\omega_m=\mathrm{const}$
direction (which is orthogonal to $\Sigma m_\nu$).

The comoving distance $D_M(z) = c \int_0^z dz'/H(z')$ accumulates
contributions from all epochs, including the low-redshift regime where the
interplay between $\Omega_m$ and $\Omega_\Lambda$ in
\begin{equation}
H(z)^2 = H_0^2 \left[\Omega_m (1+z)^3 + (1-\Omega_m)\right]
\end{equation}
breaks the degeneracy between $\Omega_m$ and $h$. At high redshift, where
$\Omega_m(1+z)^3 \gg \Omega_\Lambda$, the expansion rate simplifies to
$H(z) \approx 100\,\sqrt{\omega_m}\,(1+z)^{3/2}\;\mathrm{km\,s^{-1}\,Mpc^{-1}}$,
and distances depend on $\Omega_m$ and $h$ only through the combination
$\omega_m \equiv \Omega_m h^2$.

For the differenced observable $\Delta D_M = c\int_{z_{\rm ref}}^{z} dz'/H(z')$,
the low-redshift integral that would break this degeneracy is explicitly
removed. When both $z$ and $z_{\rm ref}$ lie in the matter-dominated regime,
\begin{equation}
\begin{gathered}
\Delta D_M \approx \frac{c}{100\,\sqrt{\omega_m}} \left[f(z) - f(z_{\rm ref})\right], \\[2pt]
f(z) \equiv 2\left[1 - (1+z)^{-1/2}\right].
\end{gathered}
\label{eq:dDM_matter}
\end{equation}
which depends only on $\omega_m$, not on $\Omega_m$ and $h$ separately.
Equation~\eqref{eq:dDM_matter} makes the structure explicit:
the differenced distance is a \emph{pure} $\omega_m$ probe (it retains the
$1/\sqrt{\omega_m}$ amplitude), and relative to the absolute distance it
loses two things. First, the $\omega_m$ amplitude is now multiplied by
$f(z)-f(z_{\rm ref})$ rather than $f(z)$, reducing its sensitivity to
$\omega_m$ (Effect~1 below). Second, having discarded the low-redshift leg,
it no longer carries the $\Omega_m$--$h$ separation that the absolute
integral obtains from the matter--dark-energy transition (Effect~2 below).

One might ask whether the expansion history near matter-radiation equality,
where $H^2(z) = (100\,\mathrm{km\,s^{-1}\,Mpc^{-1}})^2 [\omega_r(1+z)^4 + \omega_m(1+z)^3]$,
could break this degeneracy. It cannot: the physical radiation density
$\omega_r$ is fixed by $T_{\rm CMB}$, and the equality redshift
$1+z_{\rm eq} = \omega_m/\omega_r$ depends only on $\omega_m$. Measurements
of the expansion history above the matter-dark energy transition constrain
$\omega_m$ but cannot separately determine $\Omega_m$ and $h$; any
combination yielding the same product $\Omega_m h^2$ produces identical
predictions for $H(z)$. It is specifically the low-redshift regime, where
the $\Omega_m$--$\Omega_\Lambda$ interplay breaks the factorization into
$\omega_m$, that the differencing removes.

\emph{Effect 1: degradation of the $\omega_m$ amplitude
precision.} The resulting Fisher information ratio for a single tracer at
redshift $z$ is
\begin{equation}
\frac{F_{\rm diff}}{F_{\rm abs}} = \left[\frac{f(z) - f(z_{\rm ref})}{f(z)}\right]^2.
\label{eq:fisher_ratio}
\end{equation}
For $z_{\rm ref} = 0.7$ and $z = 2.0$, this gives $F_{\rm diff}/F_{\rm abs} \approx 0.20$,
or $\sigma_{\rm diff} / \sigma_{\rm abs} \approx 2.2$.
Because this derivative is taken at fixed $H_0$, varying
$\Omega_m$ is identical to varying $\omega_m = \Omega_m h^2$ up to the
constant $h^2$, which cancels in the ratio~\eqref{eq:fisher_ratio}.
Equation~\eqref{eq:fisher_ratio} is therefore a statement about
$\sigma(\omega_m)$: differencing degrades the precision on the physical
matter density by $\sim 2\times$ relative to absolute $D_M/r_d$ over the
same tracers. This is the degradation that propagates to $\Sigma m_\nu$
through $\omega_\nu = \omega_m - \omega_b - \omega_c$.
Our numerical Fisher analysis using DESI DR2 uncertainties yields a
degradation factor of approximately 2, slightly smaller because the
reference redshift $z_{\rm ref} = 0.706$ is not fully in the
matter-dominated regime (dark energy contributes $\sim$30\% of the energy
density at this epoch), so the differenced integral retains some residual
sensitivity to $\omega_m$ beyond the pure matter-dominated scaling.
The full MCMC chains corroborate this with $h$ free: in
$\Lambda$CDM the differenced observable yields $\sigma(\omega_m)$ a factor
$1.99$ larger than absolute $D_M/r_d$, and $\sigma(\Sigma m_\nu)$ a factor
$1.86$ larger, the two tracking each other as expected when
$\sigma(\Sigma m_\nu)\propto\sigma(\omega_m)$.

\begin{figure}
\centering
\includegraphics[width=\columnwidth]{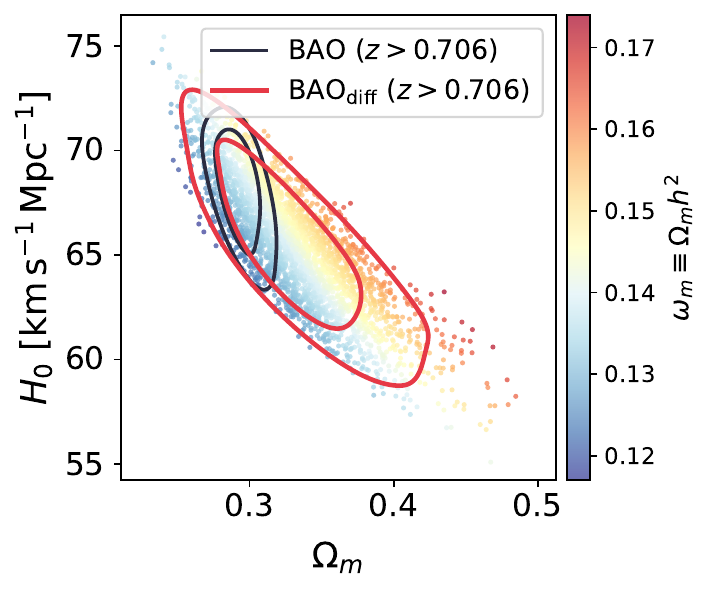}
\caption{Constraints in the $\Omega_m$--$H_0$ plane from BAO measurements at $z > 0.706$ (LRG3+ELG1 through Ly$\alpha$). Black contours ($68\%$ and $95\%$) use the absolute $D_M/r_d$ and $D_H/r_d$ measurements; red contours use the differenced approach, $\Delta D_M/r_d$ relative to LRG2 at $z_{\rm ref} = 0.706$, with $D_H/r_d$ retained. Points are samples from the differenced chain, colored by the physical matter density $\omega_m \equiv \Omega_m h^2$: the absolute constraint spans a narrow range of $\omega_m$ while the differenced constraint extends along the direction of varying $\omega_m$, showing that the differencing surrenders the matter-density anchor. The degradation arises from partial cancellation of the common $1/\sqrt{\Omega_m}$ scaling in the transverse distance differences.}

\label{fig:bao_diff_omegam}
\end{figure}

\emph{Effect 2: loss of the $\Omega_m$--$h$ degeneracy
breaking.} With $h$ free, the absolute distance additionally separates
$\Omega_m$ from $h$ through its low-redshift leg, where the
$\Omega_m$--$\Omega_\Lambda$ interplay lifts the matter-dominated
factorization into $\omega_m$. The differenced observable, having removed
that leg, cannot: it constrains $h$ only along the
$\omega_m=\mathrm{const}$ direction. Consequently $\sigma(\Omega_m)$ and
$\sigma(H_0)$ broaden by more than the factor in
Eq.~\eqref{eq:fisher_ratio} once $h$ is marginalised. This broadening is,
however, directed along $\omega_m=\mathrm{const}$, which is orthogonal to
the $\omega_m$ channel that carries $\Sigma m_\nu$; it therefore does
\emph{not} affect the neutrino-mass bound. In $\Lambda$CDM the CMB already
pins $h$ tightly, so this effect adds little beyond Effect~1
($\sigma(\Omega_m)$ degrades by $2.33\times$ versus $\sigma(\omega_m)$ by
$1.99\times$); it becomes large only when $h$ is genuinely unconstrained,
as in the BAO-only comparison of Fig.~\ref{fig:bao_diff_omegam}.

Finally, we note that the $\omega_m$ cost quantified by
Effect~1 is realised only in $\Lambda$CDM, where absolute $D_M/r_d$ pins
$\omega_m$ jointly with the CMB acoustic scale $\theta_\star$ through its
low-redshift leg. Once $(w_0,w_a)$ are freed, that joint pin is itself
removed (it is dark-energy-sensitive; Section~\ref{sec:bao}), so absolute
and differenced distances converge to nearly the same $\sigma(\omega_m)$.
This is why differencing carries no $\Sigma m_\nu$ penalty in the
late-Universe-free combination: the $\omega_m$ amplitude it discards was
DE-sensitive information, not robust information.

Figure~\ref{fig:bao_diff_omegam} confirms the picture of Effect~2:
differencing broadens the joint $(\Omega_m, H_0)$ posterior along the
$\omega_m=\mathrm{const}$ direction.

\end{document}